\documentclass[submission,copyright,creativecommons,fleqn]{eptcs}

\newcommand{\fullversion}[1]{}
\newcommand{\iceonly}[1]{}

\providecommand{\event}{ICE 2015} 

\usepackage{breakurl}             
\usepackage{hyperref}
\hypersetup{bookmarks=true,bookmarksnumbered=true,bookmarksopen}
\usepackage{amsmath,amsthm}
\usepackage[usenames]{color}\usepackage{amssymb}
\usepackage{procalg}
\usepackage{amscd}
\usepackage{txfonts}
\usepackage{amsfonts,semantic,colortbl,mathrsfs,stmaryrd,mathtools}
\usepackage{epsfig,graphicx,subfigure}
\usepackage{enumerate,enumitem}
\usepackage{multirow,multicol}
\usepackage{tabularx}
\usepackage{gastex}
\usepackage{relsize}

\newcommand{\A}{\mathcal{A}}
\newcommand{\Atau}{\mathcal{A}_{\tau}}
\newcommand{\Api}{\mathcal{A}_{\pi}}
\newcommand{\N}{\mathcal{N}}
\newcommand{\R}{\mathcal{\mathop{R}}}
\newcommand{\T}{\mathcal{T}}
\newcommand{\M}{\mathcal{M}}
\newcommand{\D}{\mathcal{D}}

\newcommand{\Dbox}{\mathcal{D}_{\Box}}
\newcommand{\Sta}{\mathcal{S}}
\newcommand{\Conf}[1][]{\mathcal{C}_{#1}}

\newcommand{\ddef}{\overset{\textrm{def}}{=}}

\newcommand{\tphdL}[1]{{#1 \!}^{\scriptscriptstyle <}}
\newcommand{\tphdR}[1]{\prescript{\scriptscriptstyle >}{}{\! #1}}

\newcommand{\nil}{\ensuremath{\mathalpha{\mathbf{0}}}}
\newcommand{\outcap}[2]{\ensuremath{\mathalpha{\overline{#1}{#2}}}}
\newcommand{\incap}[2]{\ensuremath{\mathalpha{{#1}(#2)}}}
\newcommand{\taucap}{\ensuremath{\mathalpha{\tau}}}
\newcommand{\pref}[1]{\ensuremath{\mathalpha{#1}.}}
\renewcommand{\parc}{\ensuremath{\mathbin{\mid}}}
\newcommand{\restr}[2]{\ensuremath{(#1){#2}}}
\newcommand{\repl}[1]{\ensuremath{\mathalpha{!}#1}}

\newcommand{\fn}[1]{\ensuremath{\mathsf{fn}(#1)}}
\newcommand{\bn}[1]{\ensuremath{\mathsf{bn}(#1)}}

\newcommand{\piact}[1][\alpha]{\ensuremath{\mathalpha{#1}}}
\newcommand{\inact}[2]{\ensuremath{\mathalpha{{#1}#2}}}
\newcommand{\outact}[2]{\ensuremath{\mathalpha{\overline{#1}{#2}}}}
\newcommand{\boutact}[2]{\ensuremath{\mathalpha{\overline{#1}(#2)}}}
\newcommand{\tauact}{\ensuremath{\mathalpha{\tau}}}

\newcommand{\defeqn}{\mathrel{\stackrel{\text{def}}{=}}}

\newcommand{\bbisimd}{\ensuremath{\mathrel{\bbisim^{\Delta}}}}

\renewcommand{\parc}{\ensuremath{\mathrel{\mid}}}

\newcommand{\mlarge}[1]{\mathop{\mathlarger{\mathlarger{#1}}}}
\newtheorem{definition}{Definition}
\newtheorem{lemma}{Lemma}
\newtheorem{remark}{Remark}
\newtheorem{theorem}{Theorem}
\newtheorem{corollary}{Corollary}
\newtheorem{example}{Example}

\usepackage{color}
\makeatletter
\@ifundefined{ifAP}{%

\newcommand{\todo}[1]{\underline{\sf todo}: {\color{red}{#1}}}
}{
\ifAP
\else
\fi
\newcommand{\todo}[1]{}
}
\makeatother

\newenvironment{tarray}[2][c]{
  \settowidth{\dimen1}{${} = {}$}%
  \setlength{\arraycolsep}{0.125\dimen1}%
  \hspace{-0.125\dimen1}\array[#1]{#2}\relax
}
{
   \endarray\hspace{-0.125\dimen1}
}
\def\tbb{\begin{tarray}{llll}}
\def\tbbt{\begin{tarray}[t]{llll}}
\def\tee{\end{tarray}}

\title{Executable Behaviour and the $\pi$-Calculus\\
       (extended abstract)}
\author{Bas Luttik
\institute{Department of Mathematics and Computer Science,\\ Eindhoven University of Technology\\ Eindhoven, The Netherlands}
\institute{Department of Computer Science\\ VU University Amsterdam\\ Amsterdam, The Netherlands}
\email{s.p.luttik@tue.nl}
\and
Fei Yang\thanks{This author is sponsored by the China Scholarship Council (CSC).}
\institute{Department of Mathematics and Computer Science, \\Eindhoven University of Technology\\ Eindhoven, The Netherlands}
\email{\quad f.yang@tue.nl} 
}
\def\titlerunning{Executable Behaviour and the $\pi$-Calculus}
\def\authorrunning{Bas Luttik \& Fei Yang}
\begin{document}
\maketitle

\begin{abstract}
Reactive Turing machines extend classical Turing machines with a
facility to model observable interactive behaviour. We call a
behaviour executable if, and only if, it is behaviourally equivalent
to the behaviour of a reactive Turing machine. In this paper, we study the
relationship between executable behaviour and behaviour that can be
specified in the $\pi$-calculus. We establish that all executable
behaviour can be specified in the $\pi$-calculus up to
divergence-preserving branching bisimilarity. The converse, however,
is not true due to (intended) limitations of the model of reactive
Turing machines. That is, the $\pi$-calculus allows the specification of
behaviour that is not executable up to divergence-preserving branching
bisimilarity. Motivated by an intuitive understanding of
executability, we then consider a restriction on the operational semantics
of the $\pi$-calculus that does associate with every $\pi$-term executable
behaviour, at least up to the version of branching bisimilarity that
does not require the preservation of divergence.
\end{abstract} 
\section{Introduction}\label{sec:intro}

The Turing machine~\cite{Turing1936} is generally accepted as the machine model that captures precisely which functions are algorithmically computable. As a theoretical model of the behaviour of actual computing systems, however, it has limitations, as was already observed, e.g., by Petri \cite{Pet62}. Most notably, Turing machines lack facilities to adequately deal with two important ingredients of modern computing: \emph{interaction} and \emph{non-termination}. Concurrency theory emerged from the work of Petri and developed into an active field of research. It resulted in a plethora of calculi for the formal specification of the behaviour of reactive systems, of which the $\pi$-calculus \cite{Milner1992,SW01} is probably the best-known to date.

Research in concurrency theory has focussed on defining expressive process specification formalisms, modal logics, studying suitable behavioural equivalences, etc. Expressiveness questions have also been addressed extensively in concurrency theory, especially in the context of the $\pi$-calculus (see, e.g., \cite{Gorla10,Fu2010}), but mostly pertaining to the so-called \emph{relative expressiveness} of process calculi. The absolute expressiveness of process calculi, and in particular the question as to which interactive behaviour can actually be executed by a conventional computing system, has received less attention. In this paper, we consider the expressiveness of the $\pi$-calculus with respect to the model of reactive Turing machines, proposed in \cite{BLT2013} as an orthogonal extension of classical Turing machines with a facility to model interaction in the style of concurrency theory.

Reactive Turing machines serve to define which behaviour can be executed by a computing system. Formally, we associate with every reactive Turing machine a transition system, which mathematically represents its behaviour. Then, we say that a transition system is executable if it is behaviourally equivalent to the transition system of a reactive Turing machine. Process calculi generally also have their operational semantics defined by means of transition systems. Thus, we have a method to investigate the absolute expressiveness of a process calculus, by determining to what extent transition systems specified in the calculus are executable, and by determining to what extent executable transition systems can be specified in the calculus. Note that the behavioural equivalence is a parameter of the method: if a behaviour specified in the process calculus is not executable up to some fine notion of behavioural equivalence (e.g., divergence-preserving branching bisimilarity), it may still be executable up to some coarser notion of behavioural equivalence (e.g., the divergence-insensitive variant of branching bisimilarity).  The entire spectrum of behavioural equivalences (see \cite{Glabbeek1993}) is at our disposal to draw precise conclusions. We shall use the aforementioned method to characterize the expressiveness of the $\pi$-calculus.

We shall confirm that the $\pi$-calculus is expressive: every executable behaviour can be specified in the $\pi$-calculus up to divergence-preserving branching bisimilarity \cite{Glabbeek1996,Glabbeek2009}, which is the finest behavioural equivalence discussed in van Glabbeek's seminal paper on behavioural equivalences \cite{Glabbeek1993}. Our proof explains how an arbitrary reactive Turing machine can be specified in the $\pi$-calculus. The specification consists of a  component that specifies the behaviour of the tape memory, and a component that specifies the behaviour of the finite control of the reactive Turing machine under consideration. The specification of the behaviour of the tape memory is generic and elegantly uses the link mobility feature of the $\pi$-calculus.

We also prove that the converse is not true: it is possible to specify, in the $\pi$-calculus, transition systems that are not executable up to divergence-preserving branching bisimilarity. We shall analyze the discrepancy and identify two causes.
The first cause is that the $\pi$-calculus presupposes an infinite supply of names, which is technically essential both for the way input is modelled and for the way fresh name generation is implemented. The infinite supply of names in the $\pi$-calculus gives rise to an infinite alphabet of actions. The presupposed alphabet of actions of a reactive Turing machine is, however, purposely kept finite, since allowing reactive Turing machines to have an infinite alphabet of actions arguably leads to an unrealistic model of executability. As an alternative, we shall therefore investigate the executability of $\pi$-calculus behaviour subject to name restriction, considering only the observable behaviour of a $\pi$-calculus term that refers to a finite subset of the set of names. The underlying assumption is that any realistic system will be based on a finite alphabet of input symbols.
The second cause is that, even under a finite name restriction, the transition system associated with a $\pi$-calculus term may still have unbounded branching. Transition systems with unbounded branching are not executable up to divergence-preserving branching bisimilarity, but unbounded branching behaviour can be simulated at the expense of sacrificing divergence preservation. We shall establish that, given a finite name restriction, the behaviour associated with a $\pi$-term is always executable up to (the divergence insensitive variant of) branching bisimilarity.

The paper is organized as follows. In Section~\ref{sec:def}, the basic definitions of reactive Turing machines and divergence-preserving branching bisimilarity are recapitulated, and we also recall the operational semantics of the $\pi$-calculus with replication. In Section~\ref{sec:simulation}, we prove the reactive Turing power of the $\pi$-calculus modulo divergence-preserving branching bisimilarity: a finite specification of reactive Turing machines in the $\pi$-calculus is proposed and verified. In Section~\ref{sec:exe}, we discuss the executability of transition systems associated with $\pi$-calculus processes. First, we discuss reactive Turing machines based on an infinite alphabet of actions, and argue that then, trivially, every transition system associated with a $\pi$-calculus term can be simulated up to divergence-preserving branching bisimilarity, but that the ensued notion of executability is unrealistic. Then, we establish that every finite name restriction of a behaviour specifiable in the $\pi$-calculus is executable modulo the divergence-insensitive variant of branching bisimilarity.
The paper ends with a discussion of related work and some conclusions in Section~\ref{sec:conclusions}.

\iceonly{Note to the reviewers: we put some selected proofs in the appendix. Also, the full version of this article with all detailed proofs is available as~\cite{LY14}.}

\section{A Mathematical Theory of Behaviour}\label{sec:def}

The transition system is the central notion in the mathematical theory of discrete-event behaviour. It is parameterised by a set $\A$ of \emph{action symbols}, denoting the observable events of a system. We shall later impose extra restrictions on $\A$, e.g., requiring that it be finite or have a particular structure, but for now we let $\A$ be just an arbitrary abstract set. We extend $\A$ with a special symbol $\tau$, which intuitively denotes unobservable internal activity of the system. We shall abbreviate $\A \cup\{\tau\}$ by $\Atau$.

\begin{definition}
[Labelled Transition System]\label{def:lts}
An \emph{$\Atau$-labelled transition system} $T$ is a triple $(\Sta,\step{},\uparrow)$, where,
\begin{enumerate}
    \item $\Sta$ is a set of \emph{states},
    \item ${\step{}}\subseteq\Sta\times\Atau\times\Sta$ is an $\Atau$-labelled \emph{transition relation}. If $(s,a,t)\in{\step{}}$, we write $s\step{a} t$.
    \item ${\uparrow}\in\Sta$ is the initial state.
\end{enumerate}
\end{definition}

Let $(\Sta,\step{},\uparrow)$ be an $\Atau$-labelled LTS; we define the set of reachable states from a state $s$ as follows.
\begin{equation*}
\textit{Reach}(s)=\{s'\in \Sta\mid \exists n\geq 0\,\exists s_0,\ldots,s_n\in\Sta,\, a_1,\ldots,a_n\in\Atau.\, s=s_0\step{a_1}\cdots\step{a_n}s_n=s'\}
\enskip.
\end{equation*}

Transition systems can be used to give semantics to programming languages and process calculi. The standard method is to first associate with every program or process expression a transition system (its operational semantics), and then consider programs and process expressions modulo one of the many behavioural equivalences on transition systems that have been studied in the literature. In this paper, we shall use the notion of (divergence-preserving) branching bisimilarity \cite{Glabbeek1996,Glabbeek2009}, which is the finest behavioural equivalence from van Glabbeek's linear time - branching time spectrum~\cite{Glabbeek1993}.

In the definition of (divergence-preserving) branching bisimilarity we need the following notation: let $\step{}$ be an $\Atau$-labelled transition relation on a set $\Sta$, and let $a\in\Atau$; we write $s\step{(a)}t$ for ``$s\step{a}t$ or $a=\tau$ and $s=t$''. Furthermore, we denote the transitive closure of $\step{\tau}$ by $\step{}^{+}$ and the reflexive-transitive closure of $\step{\tau}$ by $\step{}^{*}$.

\begin{definition}
[Branching Bisimilarity]\label{def:bbisim}
Let $T_1=(\Sta_1,\step{}_1,\uparrow_1)$ and $T_2=(\Sta_2,\step{}_2,\uparrow_2)$ be transition systems. A \emph{branching bisimulation} from $T_1$ to $T_2$ is a binary relation $\R\subseteq\Sta_1\times\Sta_2$ such that for all states $s_1$ and $s_2$, $s_1\R s_2$ implies
\begin{enumerate}
    \item if $s_1\step{a}_1s_1'$, then there exist $s_2',s_2''\in\Sta_2$, such that $s_2\step{}_2^{*}s_2''\step{(a)}s_2'$, $s_1\R s_2''$ and $s_1'\R s_2'$;
    \item if $s_2\step{a}_2s_2'$, then there exist $s_1',s_1''\in\Sta_1$, such that $s_1\step{}_1^{*}s_1''\step{(a)}s_1'$, $s_1''\R s_2$ and $s_1'\R s_2'$.
\end{enumerate}
The transition systems $T_1$ and $T_2$ are \emph{branching bisimilar} (notation: $T_1\bbisim T_2$) if there exists a branching bisimulation $\R$ from $T_1$ to $T_2$ s.t. $\uparrow_1\R\uparrow_2$.

A branching bisimulation $\R$ from $T_1$ to $T_2$ is \emph{divergence-preserving} if, for all states $s_1$ and $s_2$, $s_1\R s_2$ implies
\begin{enumerate}
\setcounter{enumi}{2}
    \item if there exists an infinite sequence $(s_{1,i})_{i\in\mathbb{N}}$ such that $s_1=s_{1,0},\,s_{1,i}\step{\tau}s_{1,i+1}$ and $s_{1,i}\R s_2$ for all $i\in\mathbb{N}$, then there exists a state $s_2'$ such that $s_2\step{}^{+}s_2'$ and $s_{1,i}\R s_2'$ for some $i\in\mathbb{N}$; and
    \item if there exists an infinite sequence $(s_{2,i})_{i\in\mathbb{N}}$ such that $s_2=s_{2,0},\,s_{2,i}\step{\tau}s_{2,i+1}$ and $s_1\R s_{2,i}$ for all $i\in\mathbb{N}$, then there exists a state $s_1'$ such that $s_1\step{}^{+}s_1'$ and $s_1'\R s_{2,i}$ for some $i\in\mathbb{N}$.
\end{enumerate}
The transition systems $T_1$ and $T_2$ are \emph{divergence-preserving branching bisimilar} (notation: $T_1\bbisim^{\Delta}T_2$) if there exists a divergence-preserving branching bisimulation $\R$ from $T_1$ to $T_2$ such that $\uparrow_1\R\uparrow_2$.
\end{definition}

For two LTSs $T_1=(\Sta_1,\step{}_1,\uparrow_1)$ and $T_2=(\Sta_2,\step{}_2,\uparrow_2)$, $s_1\in\Sta_1$ and $s_2\in\Sta_2$, we write $s_1\bbisim s_2$ ($s_1\bbisimd s_2$) if there is a (divergence-preserving) branching bisimilarity from $T_1$ to $T_2$ relating $s_1$ and $s_2$. Thus, $\bbisim$ is a relation from the states of $T_1$ to the states of $T_2$, and it can be shown that it satisfies the conditions of Definition~\ref{def:bbisim}.
We can also write $s_1\bbisim s_2$ ($s_1\bbisimd s_2$) if $s_1$ and $s_2$ are states in a single LTS $T$ and related by a (divergence-preserving) branching bisimulation from $T$ to itself.

The relations $\bbisim$ and $\bbisimd$ are equivalence relations, both as relations on a single transition system, and as relations on a set of transition systems \cite{Bas96,Glabbeek2009}.

Next we define the notion of bisimulation up to $\bbisim$. Note that we adapt a non-symmetric bisimulation up to relation, which is a useful tool to establish the proofs of $\bbisim$ later.

\begin{definition}\label{def:up-to}
Let $T_1=(\Sta_1,\step{}_1,\uparrow_1)$ and $T_2=(\Sta_2,\step{}_2,\uparrow_2)$ be two transition systems. A relation $\R\subseteq\Sta_1\times\Sta_2$ is a bisimulation up to $\bbisim$ if, whenever $s_1\R s_2$, then for all $a\in \Atau$:
\begin{enumerate}
    \item if $s_1\step{}^{*}s_1''\step{a}s_1'$, with $s_1\bbisim s_1''$ and ${a\neq\tau}\vee{s_1''\not\bbisim s_1'}$, then there exists $s_2'$ such that $s_2\step{a}s_2'$, $s_1''\mathrel{\bbisim\mathrel{\circ}\mathrel{\R}}s_2$ and $s_1'\mathrel{\bbisim \mathrel{\circ} \mathrel{\R}} s_2'$; and
    \item if $s_2\step{a}s_2'$, then there exist $s_1',s_1''$ such that $s_1\step{}^{*}s_1''\step{a}s_1'$, $s_1''\bbisim s_1$ and $s_1'\mathrel{\bbisim \mathrel{\circ} \mathrel{\R}} s_2'$.
\end{enumerate}
\end{definition}

\begin{lemma}\label{lemma:up-to}
If $\R$ is a bisimulation up to $\bbisim$, then $\R \subseteq {\bbisim}$.
\end{lemma}
\fullversion{%
\begin{proof}
It is sufficient to prove that $\mathrel{\bbisim \mathrel{\circ} \mathrel{\R}}$ is a branching bisimulation, since $\bbisim$ is a transitive relation.
Let $s_1\bbisim s_2\mathrel{\R} s_3$.
\begin{enumerate}
    \item Suppose $s_1\step{a}s_1'$. We distinguish with two cases,
    \begin{enumerate}
        \item If $a=\tau$ and $s_1\bbisim s_1'$, then $s_1'\bbisim s_1\bbisim s_2$, so $s_1'\mathrel{\bbisim \mathrel{\circ} \mathrel{\R}}s_3$.
        \item Otherwise, we have ${a\neq\tau}\vee{s_1\not\bbisim s_1'}$. Then according to Definition~\ref{def:bbisim}, there exist $s_2''$ and $s_2'$ such that $s_2\step{}^{*}s_2''\step{a}s_2'$, $s_1\bbisim s_2''$ and $s_1'\bbisim s_2'$. Note that $s_2\bbisim s_1 \bbisim s_2''$, so by Definition~\ref{def:up-to}, there exist $s_4''$, $s_4'$ and $s_3'$ such that $s_3\step{a}s_3'$ and $s_2''\bbisim s_4'' \mathrel{\R} s_3$ and $s_2'\bbisim s_4'\mathrel{\R} s_3'$. Since $s_1'\bbisim s_2'\bbisim s_4'$ and $s_4' \mathrel{\R} s_3' $, it follows that $s_1'\mathrel{\bbisim \mathrel{\circ} \mathrel{\R}} s_3'$.

    \end{enumerate}
    \item If $s_3\step{a}s_3'$, then according to Definition~\ref{def:up-to}, there exist $s_2''$ and $s_2'$ such that $s_2\step{}^{*}s_2''\step{a}s_2'$, $s_2''\bbisim s_2$ and $s_2'\mathrel{\bbisim \mathrel{\circ} \mathrel{\R}}s_3'$ , since $s_1\bbisim s_2\bbisim s_2''$ and $s_2''\step{a}s_2'$, by Definition~\ref{def:bbisim}, there exist $s_1''$ and $s_1'$ such that $s_1\step{}^{*}s_1''\step{(a)}s_1'$ with $s_1'' \bbisim s_2''$ and $s_1'\bbisim s_2'$. Since $s_2''\bbisim t\mathrel{\R} s_3$ and $s_2'\mathrel{\bbisim \mathrel{\circ} \mathrel{\R}}s_3'$, it follows that $s_1''\mathrel{\bbisim \mathrel{\circ} \mathrel{\R}}s_3$ and $s_1'\mathrel{\bbisim \mathrel{\circ} \mathrel{\R}}s_3'$.
\end{enumerate}
Therefore, a branching bisimulation up to $\bbisim$ is included in $\bbisim$.
%
\end{proof}}
\subsection{Executable behaviour}

The notion of reactive Turing machine (RTM) was put forward in \cite{BLT2013} to mathematically characterise which behaviour is executable by a conventional computing system. In this section, we recall the definition of RTMs and the ensued notion of executable transition system. The definition of RTMs is parameterised with the set $\Atau$, which we  now assume to be a finite set. Furthermore, the definition is parameterised with another finite set $\D$ of \emph{data symbols}. We extend $\D$ with a special symbol $\Box\notin\D$ to denote a blank tape cell, and denote the set $\D\cup\{\Box\}$ of \emph{tape symbols} by $\Dbox$.
\begin{definition}
[Reactive Turing Machine]\label{def:rtm}
A \emph{reactive Turing machine} (RTM) $\M$ is a triple $(\Sta,\step{},\uparrow)$, where
\begin{enumerate}
    \item $\Sta$ is a finite set of \emph{states},
    \item ${\step{}}\subseteq \Sta\times\Dbox\times\Atau\times\Dbox\times\{L,R\}\times\Sta$ is a finite collection of $(\Dbox\times\Atau\times\Dbox\times\{L,R\})$-labelled \emph{transition rules} (we write $s\step{a[d/e]M}t$ for $(s,d,a,e,M,t)\in{\step{}}$),
    \item ${\uparrow}\in\Sta$ is a distinguished \emph{initial state}.
\end{enumerate}
\end{definition}
\begin{remark}
  The original definition of RTMs in \cite{BLT2013} includes an extra facility to declare a subset of the states of an RTM as being final states, and so does the associated notion of executable transition system. In this paper, however, our goal is to explore the relationship  between the transition systems associated with RTMs and those that can be specified in the $\pi$-calculus. Since the $\pi$-calculus does not include the facility to specify that a state has the option to terminate, we leave it out from the definition of RTMs too.
\end{remark}

Intuitively, the meaning of  a transition $s\step{a[d/e]M}t$ is that whenever $\M$ is in state $s$, and $d$ is the symbol currently read by the tape head, then it may execute the action $a$, write symbol $e$ on the tape (replacing $d$), move the read/write head one position to the left or the right on the tape (depending on whether $M=L$ or $M=R$), and then end up in state $t$.

To formalise the intuitive understanding of the operational behaviour of RTMs, we associate with every RTM $\M$ an $\Atau$-labelled transition system  $\T(\M)$. The states of $\T(\M)$ are the
configurations of $\M$, which consist of a state from $\Sta$, its tape contents, and the position of the read/write head. 
We denote by $\check{\Dbox}=\{\check{d}\mid d\in\Dbox\}$ the set of \emph{marked} symbols; a \emph{tape instance} is a sequence $\delta\in(\Dbox\cup\check{\Dbox})^{*}$ such that $\delta$ contains exactly one element of $\check{\Dbox}$, indicating the position of the read/write head.
We adopt a convention to concisely denote new placement of the tape head marker. Let $\delta$ be an element of $\Dbox^{*}$. Then by $\tphdL{\delta}$ we denote the element of $(\Dbox\cup\check{\Dbox})^{*}$ obtained by placing the tape head marker on the right-most symbol of $\delta$ (if it exists), and $\check{\Box}$ otherwise.
Similarly $\tphdR{\delta}$ is obtained by placing the tape head marker on the left-most symbol of $\delta$ (if it exists), and $\check{\Box}$ otherwise.

\begin{definition}\label{def:lts-tm}
Let $\M=(\Sta,\step{},\uparrow)$ be an RTM. The \emph{transition system} $\T(\M)$ \emph{associated with} $\M$ is defined as follows:
\begin{enumerate}
\item its set of states is the set $\Conf[\M]=\{(s,\delta)\mid s\in\Sta,\ \text{$\delta$ a tape instance}\}$ of all configurations of $\M$;
    \item its transition relation ${\step{}}\subseteq{\Conf[\M]\times\Atau\times\Conf[\M]}$ is the least relation satisfying, for all $a\in\Atau,\,d,e\in\Dbox$ and $\delta_L,\delta_R\in\Dbox^{*}$:
    \begin{itemize}
        \item $(s,\delta_L\check{d}\delta_R)\step{a}(t,\tphdL{\delta_L}e\delta_R)$ iff $s\step{a[d/e]L}t$, and
        \item $(s,\delta_L\check{d}\delta_R)\step{a}(t,\delta_L e{}\tphdR{\delta_R})$ iff $s\step{a[d/e]R}t$, and
    \end{itemize}
    \item its initial state is the configuration $(\uparrow,\check{\Box})$.
\end{enumerate}
\end{definition}

Turing introduced his machines to define the notion of \emph{effectively computable function}. By analogy, the notion of RTM can be used to define a notion of \emph{effectively executable behaviour}.

\begin{definition}
[Executability]\label{def:exe}
A transition system is \emph{executable} if it is the transition system associated with some RTM.
\end{definition}

Usually, we shall be interested in executability up to some behavioural equivalence. In \cite{BLT2013}, a characterisation of executability up to (divergence-preserving) branching bisimilarity is given that is independent of the notion of RTM.
In order to be able to recapitulate the results below, we need the following definitions, pertaining to the recursive complexity and branching degree of transition systems.
Let $T=(\Sta,\step{},\uparrow)$ be a transition system. We say that $T$ is \emph{effective} if $\step{}$ is a recursively enumerable set. The mapping $\mathalpha{out}:\Sta\rightarrow 2^{\A_{\tau}\times\Sta}$ associates with every state its set of outgoing transitions, i.e., for all $s\in\Sta$, $\mathalpha{out}(s)=\{(a,t)\mid s\step{a}t\}$. We say that $T$ is \emph{computable} if $\mathalpha{out}$ is a recursive function. We call a transition system \emph{finitely branching} if $\mathalpha{out}(s)$ is finite for every state $s$, and \emph{boundedly branching} if there exists $B\in\mathbb{N}$ such that $|\mathalpha{out}(s)|\leq B$ for all $s\in \Sta$.

The following results were established in \cite{BLT2013}.
\begin{theorem}\label{thm-blt}
\begin{enumerate}
\item
  The transition system $\T(\M)$ associated with an RTM $\M$ is computable and boundedly branching.
\item
  For every finite set $\Atau$ and every boundedly branching computable $\Atau$-labelled transition system $T$, there exists an RTM $\M$ such that $T\bbisim^{\Delta} \T(\M)$.
\item
  For every finite set  $\Atau$ and every effective $\Atau$-labelled transition system $T$ there exists an RTM $\M$ such that $T\bbisim \T(\M)$.
\end{enumerate}
\end{theorem}

Notice the role played by divergence preservation in the preceding theorem. Divergence can be used to simulate the behaviour in a state with a high branching degree using states with lower branching degrees; the idea stems from \cite{Baeten1987129} and is generalised in \cite{Phi93} to prove that every effective $\Atau$-labelled transition system is weakly bisimilar to a computable transition system. We proceed to discuss a criterion to decide whether a transition system $T=(\Sta,\step{},\uparrow)$ is not executable up to divergence-preserving branching bisimilarity, which is based on the notion of branching degree up to $\bbisimd$. Let us denote the equivalence class of $s\in\Sta$ modulo $\bbisimd$ by $[s]_{\bbisimd}=\{s'\in\Sta\mid s\bbisimd s'\}$.
The \emph{branching degree} up to $\bbisimd$ of $s$, denoted by $deg_{\bbisimd}(s)$, is defined as the cardinality of the set
\begin{equation*}
  \mlarge{\{} {(a,[s']_{\bbisimd})} \mlarge{\mid} \exists s'' .\, s\step{}^{*} s''\step{a}s' \ \&\ s\bbisimd s''\ \&\ (a=\tau \implies s''\not\bbisimd s') \mlarge{\}}
\enskip.
\end{equation*}
The branching degree modulo $\bbisimd$ of $T$ is the least upper bound of the branching degrees of all reachable states, which is defined as $deg_{\bbisimd}(T)=\sup\{deg_{\bbisimd}(s)\mid s\in \textit{Reach}(\uparrow)\}$. We say that $T$ is \emph{boundedly branching} up to $\bbisimd$ if there exists $B\in\mathbb{N}$, such that $deg_{\bbisimd}(T)\leq B$, otherwise it is \emph{unboundedly branching} up to $\bbisimd$.

\begin{lemma} \label{lemma:bbisimddegpres}
  If $s\bbisimd t$, then $deg_{\bbisimd}(s) = deg_{\bbisimd}(t)$.
\end{lemma}
\fullversion{%
\begin{proof}
  Clearly, if there exist $s''$ and $s'$ such that $s \step{}^{*} s'' \step{a} s'$, then it is straightforward to derive from the definition of $\bbisimd$ that there also exist $t''$ and $t'$ such that $t \step{}^{*} t'' \step{a} t'$ and $s''\bbisimd t''$. Conversely, if there exist $t''$ and $t'$ such that $t \step{}^{*} t'' \step{a} t'$, then there also exist $s''$ and $s'$ such that $s \step{}^{*} s'' \step{a} s'$ and $s''\bbisimd t''$. Hence, there is a bijective correspondence between the sets
\begin{equation*}
  \mlarge{\{}{(a,[s']_{\bbisimd})} \mlarge{\mid} \exists s'' .\, s\step{}^{*} s''\step{a}s' \ \&\ s\bbisimd s''\ \&\ (a=\tau \implies s''\not\bbisimd s')\mlarge{\}}
\end{equation*}
and
\begin{equation*}
  \mlarge{\{}{(a,[t']_{\bbisimd})} \mlarge{\mid} \exists t'' .\, t\step{}^{*} t''\step{a}t' \ \&\ t\bbisimd t''\ \&\ (a=\tau \implies t''\not\bbisimd t')\mlarge{\}}
\enskip.
\end{equation*}
It follows that $deg_{\bbisimd}(s) = deg_{\bbisimd}(t)$.
\end{proof}}

A \emph{divergence} (up to $\bbisim^{\Delta}$) in a transition system is an infinite sequence of reachable states $s_1,s_2,\dots$ such that $s_{1}\step{\tau}s_{2}\step{\tau}\cdots$ and $s_{i}\bbisim^{\Delta} s_{j}$ for all $i,j\in\mathbb{N}$. The following lemma shows that, in the absence of a divergence, boundedly branching transition systems are boundedly branching up to $\bbisimd$.

\begin{lemma}\label{lemma:divergence}
If a transition system is boundedly branching and does not have divergence up to $\bbisimd$, then it is boundedly branching up to $\bbisimd$.
\end{lemma}
\fullversion{%
\begin{proof}
Let $T$ be a boundedly branching transition system and suppose that $T$ does not have a divergence up to $\bbisimd$.
Then there exists $B\in\mathbb{N}$ such that $|\mathalpha{out}(s)|\leq B$ for all states $s$ of $T$. It suffices to prove that $deg_{\bbisimd}(s)\leq B$ for all states $s$ of $T$. To this end, let $s$ be a state of $T$. Since $T$ does not have divergence up to $\bbisimd$, there exists $t$ such that $s\step{}^{*}t$ and $s\bbisimd t$ and there does not exist $t'$ such that $t\step{\tau}t'$ and $t\not\bbisimd t'$. Then
\begin{equation*}
   \{ \{{(a,[t']_{\bbisimd})} \mid \exists t'' .\, t\step{}^{*} t''\step{a}t' \ \&\ t\bbisimd t''\ \&\ (a=\tau \implies t''\not\bbisimd t')\}
    =
  \{{(a,[t']_{\bbisimd})} \mid t\step{a}t' \}
\enskip.
\end{equation*}
From $s \bbisimd t$ it follows by Lemma~\ref{lemma:bbisimddegpres} that $deg_{\bbisimd}(s) =deg_{\bbisimd}(t)$, so
\begin{equation*}
  deg_{\bbisimd}(s)
    =deg_{\bbisimd}(t)
    =|\{{(a,[t']_{\bbisimd})} \mid t\step{a}t' \}|
    \leq |\{(a,t') \mid t\step{a}t' \}|
    =|\mathalpha{out}(t)|
    \leq B
\enskip.
\end{equation*}
We conclude that $T$ is boundedly branching up to $\bbisimd$.
\end{proof}}

Thus we conclude the following theorem from Theorem~\ref{thm-blt}(1) and Lemma~\ref{lemma:divergence}.

\begin{theorem}\label{thm:non-exe}
If a transition system $T$ has no divergence up to $\bbisimd$ and is unboundedly branching up to $\bbisimd$, then it is not executable modulo $\bbisimd$.
\end{theorem}

\subsection{$\pi$-Calculus}\label{sec:pi}

The $\pi$-calculus was proposed by Milner, Parrow and Walker in~\cite{Milner1992} as a language to specify processes with link mobility. The expressiveness of many variants of the $\pi$-calculus has been extensively studied. In this paper, we shall consider the basic version presented in \cite{SW01}, excluding the match prefix. We recapitulate some definitions from \cite{SW01} below and refer to the book for detailed explanations.

We presuppose a countably infinite set $\N$ of names; we use strings of lower case letters for elements of $\N$.
The \emph{prefixes}, \emph{processes} and \emph{summations} of the $\pi$-calculus are, respectively, defined by the following grammar:
\begin{align*}
\pi\      & \coloneqq\ \outcap{x}{y}\ \mid\ \incap{x}{z}\ \mid\ \taucap \qquad (x,y,z\in \N)\\
P\    & \coloneqq\ M\ \mid\  P\parc P\ \mid\ \restr{z}{P}\ \mid\ \repl{P}\\
M\   & \coloneqq\ \nil\ \mid\ \pref{\pi}P \mid\ M \altc M\enskip.
\end{align*}

In $\pref{\incap{x}{z}}P$ and $\restr{z}{P}$, the displayed occurrence of the name $z$ is \emph{binding} with scope $P$. An occurrence of a name in a process is \emph{bound} if it is, or lies within the scope of, a binding occurrence in $P$; otherwise it is free. We use $\fn{P}$ to denote the set of names that occur free in $P$, and $\bn{P}$ to denote the set of names that occur bound in $P$.

An $\alpha$-conversion between $\pi$-terms is defined in~\cite{SW01} as a finite number of changes of bound names. In this paper, we do not distinguish among $\pi$-terms that are $\alpha$-convertible, and we write $P=Q$ if $P$ and $Q$ are $\alpha$-convertible.

We define the operational behaviour of $\pi$-processes by means of the structural operational semantics in Fig.~\ref{tab:pi-semantics}, in which $\piact{}$ ranges over the set of actions of the $\pi$-calculus
\begin{equation*}
  \Api =\{\inact{x}{y},\outact{x}{y},\boutact{x}{z}\mid x,y,z \in\N\}\cup\{\tau\}
\enskip.
\end{equation*}
\begin{figure}

\begin{center}
\fbox{
\begin{minipage}[t]{0.9\textwidth}
$\mathrm{PREFIX}\quad\inference{\,}{\tau.P\step{\tau}P}\quad \inference{}{\overline{x}y.P\step{\overline{x}y}P}\quad\inference{}{x(y).P\step{xz}P\{z/y\}}$\\
$\mathrm{SUM_L}\quad\inference{P\step{\piact{}}P'}{(P+Q)\step{\piact{}}P'}
\quad\mathrm{PAR_L}\quad\inference{P\step{\piact{}}P'}{P\parc{Q}\step{\piact{}}P'\parc{Q}}\,\bn{\piact{}}\cap \fn{Q}=\emptyset$\\
$\mathrm{COM_L}\quad\inference{P\step{\overline{x}y}P',\,Q\step{xy}Q'}{P\parc{Q}\step{\tau}P'\parc{Q'}}\quad
\mathrm{CLOSE_L}\quad\inference{P\step{\overline{x}(z)}P',\,Q\step{xz}Q'}{P\parc{Q}\step{\tau}(z)(P'\parc{Q'})}\,z\notin \fn{Q}$\\
$\mathrm{RES}\quad\inference{P\step{\piact{}}P'}{\restr{z}{P}\step{\piact{}}\restr{z}{P'}}\,z\notin\piact{}\quad
\mathrm{OPEN}\quad\inference{P\step{\overline{x}z}P'}{(z)P\step{\overline{x}(z)}P'}\,z\neq x$\\
$\mathrm{REP}
  \quad\inference{P\step{\piact{}}P'}{\repl{P}\step{\piact{}}P'\parc\repl{P}}
  \quad\inference{P\step{\outact{x}{y}}P',\,P\step{\inact{x}{y}}P''}{\repl{P}\step{\tauact{}}(P'\parc P'')\parc\repl{P}}
  \quad\inference{P\step{\boutact{x}{z}}P',\,P\step{\inact{x}{z}}P''}{\repl{P}\step{\tauact}\restr{z}{(P'\parc P'')}\parc\repl{P}}$
\end{minipage}
}\end{center}
\caption{Operational rules for the $\pi$-calculus}\label{tab:pi-semantics}
\end{figure}

The rules in Fig.~\ref{tab:pi-semantics} define on $\pi$-terms an $\Api$-labelled transition relation ${\step{}}$.
Then, we can associate with every $\pi$-term $P$ an $\Api$-labelled transition system $\T(P)=(\Sta_{P},\step{}_{P},P)$. The set of states $\Sta_{P}$ of $\T(P)$ consists of all $\pi$-terms reachable from $P$, the transition relation $\step{}_{P}$ of $\T(P)$ is obtained by restricting the transition relation $\step{}$ defined by the structural operational rules to $\Sta_{P}$ (i.e., ${\step{}_{P}}={\step{}}\cap (\Sta_{P}\times \Api \times\Sta_{P})$), and the initial state of $\T(P)$ is the $\pi$-term $P$.



For convenience, we sometimes want to abbreviate interactions that involve the transmission of no name at all, or more than one name. Instead of giving a full treatment of the polyadic $\pi$-calculus (see \cite{SW01}), we define the following abbreviations:
\begin{align*}
   \pref{\outcap{x}{\langle y_1,\dots,y_n\rangle}}P
      &\defeqn\restr{w}{\pref{\outcap{x}{w}}\pref{\outcap{w}{y_1}}\cdots\pref{\outcap{w}{y_n}}P}\quad (w\not\in\fn{P}),\ \text{and}\\
   \pref{\incap{x}{z_1,\dots,z_n}}P
       &\defeqn\pref{\incap{x}{w}}\pref{\incap{w}{z_1}}\cdots\pref{\incap{w}{z_n}}P
 \enskip.
\end{align*}

The following lemma establishes that divergence-preserving branching bisimilarity is compatible with restriction and parallel composition. This will be a useful property when establishing the correctness of our simulation of RTMs in the $\pi$-calculus, in the next section.

\begin{lemma}\label{lemma:pi-compat}
For all $\pi$-terms $P$, $P'$, $Q$, and $Q'$:
\begin{enumerate}
\item\label{pi-compat:restriction} if $P\bbisimd P'$, then $(a)P \bbisimd (a)P'$;
\item if $P\bbisimd P'$ and $Q\bbisimd Q'$, then $P\parc{Q}\bbisimd P'\parc{Q'}$.
\end{enumerate}
\end{lemma}
\fullversion{%
\begin{proof}
\begin{enumerate}
    \item It is straightforward to verify that the relation
       \begin{equation*}
          \R = \{((a)P,(a)P')\mid \text{$P$ and $P'$ are $\pi$-terms s.t.\ $P\bbisimd{}P'$}\}
       \end{equation*}
       is a divergence-preserving branching bisimulation relation.
    \item Define the relation $\R$ by
       \begin{multline*}
          \{((\vec{a})(P\parc{} Q), (\vec{a})(P'\parc{} Q'))\mid
 \text{$\vec{a}$ is a sequence of restricted names and}\\
\text{$P$, $P'$, $Q$, and $Q'$ are $\pi$-terms s.t.\ $P\bbisimd{} P'$ and $Q\bbisimd{} Q'$}\}
       \enskip;
       \end{multline*}
       we verify that $\R$ is a divergence-preserving branching bisimulation. For simplicity, we only analyze the terms with no restricted names.
       To this end, we first suppose that $P\parc{} Q\step{a}R$; and we distinguish three cases according to which operational rule is applied last in the derivation of this transition:
        \begin{enumerate}
            \item if $P\step{a}P_1$ and $R = P_1\parc{}Q$, then, since $P\bbisimd{}P'$, there exist $P_1''$ and $P_1'$, such that $P'\step{}^{*}P_1''\step{a} P_1'$ with $P'\bbisimd{} P_1''$ and $P_1\bbisimd{} P_1'$. Thus $P'\parc{} Q'\step{}^{*}P_1''\parc{} Q\step{a}P_1'\parc{} Q$, and, according to the definition of $\R$, we have ${P\parc{} Q}\mathrel{\R} {P_1''\parc{} Q}$ and ${P_1\parc{} Q}\mathrel{\R}{P_1'\parc{} Q}$.
            \item If $P\step{\overline{x}y}P_1$, $Q\step{xy}Q_1$, $R = P_1\parc{} Q_1$, and $a=\tau$, then, since $P\bbisim^{\Delta}P'$ and $Q\bbisimd{}Q'$, there exist $P_1''$, $P_1'$, $Q_1''$ and $Q_1'$ such that $P'\step{}^{*}P_1''\step{\overline{x}y} P_1'$ with $P\bbisimd{} P_1''$ and $P_1\bbisimd{} P_1'$, and $Q'\step{}^{*}Q_1''\step{xy} Q_1'$ with $Q\bbisimd{} Q_1''$ and $Q_1\bbisimd{} Q_1'$. Hence, it follows that
                         ${P'\parc{} Q'}\step{}^{*} P_1''\parc{} Q_1''\step{\tau} P_1'\parc{} Q_1'$, with $P\parc Q\mathrel{\R} P_1''\parc{} Q_1''$ and ${P_1\parc{} Q_1} \mathrel{\R} P_1'\parc{} Q_1'$.
            \item If $P\step{\overline{x}(z)}P_1$, $Q\step{xz}Q_1$, $R=(z)(P_1\parc{} Q_1)$, and $a=\tau$, then, since $P\bbisim^{\Delta}P'$ and $Q\bbisimd{}Q'$, there exist $P_1''$, $P_1'$, $Q_1''$ and $Q_1'$ such that $P'\step{}^{*}P_1''\step{\overline{x}(z)} P_1'$ with $P\bbisimd{} P_1''$ and $P_1\bbisimd{} P_1'$, and $Q'\step{}^{*}Q_1''\step{xz} Q_1'$ with $Q\bbisimd{} Q_1''$ and $Q_1\bbisimd{} Q_1'$. Hence, it follows that
                ${P'\parc{} Q'}\step{}^{*} P_1''\parc{} Q_1''\step{\tau} (z)(P_1'\parc{} Q_1')$, with $P\parc Q\mathrel{\R} P_1''\parc{} Q_1''$ and ${(z)(P_1\parc{} Q_1)} \mathrel{\R} (z)(P_1'\parc{} Q_1')$.
        \end{enumerate}
        The symmetric cases can be proved analogously.
\end{enumerate}
\end{proof} } 
\section{Specifying Executable Behaviour in the $\pi$-Calculus}\label{sec:simulation}


In the previous section, we have introduced the $\pi$-calculus as a language to specify behaviour of systems with link mobility, and we have proposed RTMs to define a notion of executable behaviour. In this section we prove that every executable behaviour can be specified in the $\pi$-calculus up to divergence-preserving branching bisimilarity. To this end, we associate with every RTM $\M$ a $\pi$-term $P$ that simulates the behaviour of $\M$ up to divergence-preserving branching bisimilarity, that is, $\T(\M)\bbisim^{\Delta}\T(P)$.

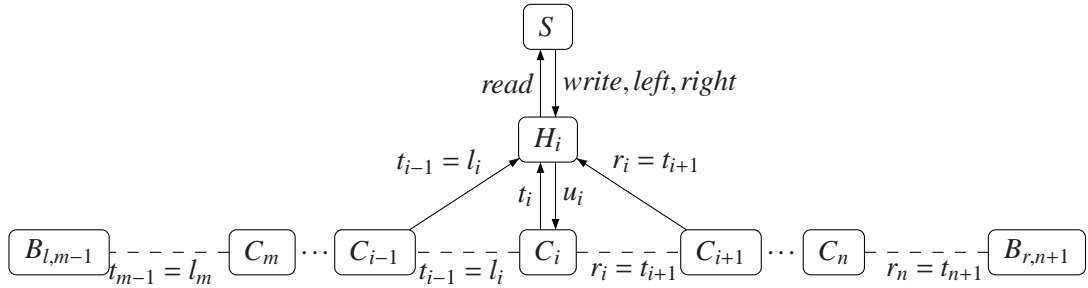
\begin{figure}
\begin{center}
\begin{picture}(80,35)(-40,-5)
  \gasset{Nadjust=w,Nadjustdist=2,Nh=6,Nmr=1}
  \node(A)(0,0){$C_i$}
  \node(B)(23,0){$C_{i+1}$}
  \node(C)(38,0){$C_n$}
  \node(D)(65,0){$B_{r,n+1}$}
  \node(E)(-23,0){$C_{i-1}$}
  \node(F)(-38,0){$C_{m}$}
  \node(G)(-65,0){$B_{l,m-1}$}
  \node(H)(0,15){$H_i$}
  \node[Nframe=n](I)(31,0){$\ldots$}
  \node[Nframe=n](J)(-31,0){$\ldots$}
  \node(K)(0,30){$S$}

  \drawedge[sxo=1,exo=1](H,A){$u_i$}
  \drawedge[sxo=-1,exo=-1](A,H){$t_i$}
  \drawedge[ELside=r](B,H){$r_i=t_{i+1}$}
  \drawedge(E,H){$t_{i-1}=l_i$}
  \drawedge[AHnb=0,dash={1.5}0,ELside=r](A,B){$r_i=t_{i+1}$}
  \drawedge[AHnb=0,dash={1.5}0](A,E){$t_{i-1}=l_i$}
  \drawedge[AHnb=0,dash={1.5}0,ELside=r](C,D){$r_{n}=t_{n+1}$}
  \drawedge[AHnb=0,dash={1.5}0](F,G){$t_{m-1}=l_{m}$}
  \drawedge[sxo=1,exo=1](K,H){$\textit{write}, \textit{left}, \textit{right}$}
  \drawedge[sxo=-1,exo=-1](H,K){$\textit{read}$}
\end{picture}
\caption{\label{fig:specification}Specification of an RTM utilizing the linking structure of the $\pi$-calculus}
\end{center}
\end{figure}

The structure of our specification is illustrated in Figure~\ref{fig:specification}. In this figure, each node represents a parallel component of the specification, each labelled arrow stands for a communication channel with certain labels, and the dashed lines represent the links between cells. Moreover, the equalities on arrows and dashed lines tell the correspondence between the names defined in the linked terms. The specification consists of a generic finite specification of the behaviour of a tape (parallel components $H_k$, $B_{l,k}$, $C_k$, $B_{r,k}$ in Figure~\ref{fig:specification}), and a finite specification of a control process that is specific for the RTM $\M$ under consideration (parallel component $S$ in Figure~\ref{fig:specification}). We first discuss the generic specification of the tape in Section~\ref{subsec:tape}, then we discuss how to add a suitable control process specific for $\M$ in Section~\ref{subsec:fincontrol} proving that $\M$ is simulated by the parallel composition of the two parts.

\subsection{Tape} \label{subsec:tape}

In~\cite{Baeten1987129}, the behaviour of the tape of a Turing machine is finitely specified in ACP$_{\tau}$ making use of finite specifications of two stacks. The specification is not easily modified to take intermediate termination into account, and therefore, in~\cite{BLT2013}, an alternative solution is presented, specifying the behaviour of a tape in TCP$_{\tau}$ by using a finite specification of a queue (see also \cite{baeten2010}). In this paper, we will exploit the link passing feature of the $\pi$-calculus to give a more direct specification. In particular, we shall model the tape as a collection of cells endowed with a link structure that organises them in a linear fashion.

We first give an informal description of the behaviour of a tape. The state of a tape is characterised by a tape instance $\delta_L\check{d}\delta_R$, consisting of a finite (but unbounded) sequence of data with the current position of the tape head indicated by ~$\check{}$~. The tape may then exhibit the following observable actions:
\begin{enumerate}
    \item $\overline{\textit{read}}\,d$: the datum under the tape head is output along the channel $\textit{read}$;
    \item $\textit{write}(e)$: a datum $e$ is written on the position of the tape head, resulting in a new tape instance $\delta_L\check{e}\delta_R$; and
    \item $\textit{left}$, $\textit{right}$:  the tape head moves one position left or right, resulting in ${\tphdL{\delta_L}}d\delta_R$ or $\delta_L d \tphdR{\delta_R}$, respectively.
\end{enumerate}

Henceforth, we assume that tape symbols are included in the set of names, i.e., that $\Dbox\subseteq\N$.

In our $\pi$-calculus specification of a tape, each individual tape cell is specified as a separate component, and there is a separate component modelling the tape head. A tape cell stores a datum $d$, represented by a free name in the specification, and it has pointers $l$ and $r$ to its left and right neighbour cells. Furthermore, it has two links to  the component modelling the tape head: the link $u$ is used by the tape head for updating the datum, and the link $t$ serves as a general communication channel for communicating all relevant information about the cell to the tape head. The following $\pi$-term represents the behaviour of a tape cell:
\begin{eqnarray*}
C &\ddef& c(t,l,r,u,d).C(t,l,r,u,d)\\
C(t,l,r,u,d) &\ddef& u(e).\overline{c}\langle t,l,r,u,e\rangle.\nil+\overline{t}\langle l,r,u,d\rangle.\overline{c}\langle t,l,r,u,d\rangle.\nil
\enskip.
\end{eqnarray*}

Note that the behaviour of an individual tape cell $C(t,l,r,u,d)$ is as follows: either it receives along channel $u$ an update $e$ for its datum $d$, after which it recreates itself with datum $e$ in place of $d$; or it outputs all relevant information about itself (i.e., the links to its left and right neighbours, its update channel $u$, and
the stored datum $d$) to the tape head along channel $t$, after which it recreates itself. A cell is created by a synchronisation on name $c$, by which all relevant information about the cell is passed; we shall have a component $!C$ facilitate the generation of new incarnations of existing tape cells.

At any moment, the number of tape cells will be finite.  To model the unbounded nature of the tape, we define a process $B$ that serves to generate new blank tape cells on either side of the  tape whenever needed:
\begin{eqnarray*}
B &\ddef&b_l(t,r).(u,l)B_l(t,l,r,u) + b_r(t,l).(u,r)B_r(t,l,r,u)\\
B_l(t,l,r,u) &\ddef& \overline{t}\langle l,r,u,\Box\rangle.(\overline{c}\langle t,l,r,u,\Box\rangle.\nil\parc{\overline{b_l}\langle l,t\rangle.\nil}) \\
B_r(t,l,r,u) &\ddef& \overline{t}\langle l,r,u,\Box\rangle.(\overline{c}\langle t,l,r,u,\Box\rangle.\nil\parc{\overline{b_r}\langle t,r\rangle.\nil})
\enskip.
\end{eqnarray*}

Note that $B$ offers the choice to either create a blank tape cell at the left-hand side of the tape through $B_l(t,l,r,u)$, or a blank tape cell at the right-hand side of the tape through $B_r(t,l,r,u)$. In the first case, suppose the original leftmost cell has the channels $t_o$ and $l_o$, for itself and its left neighbour, respectively, then for the new cell, we have $t=l_o$ and $r=t_o$, in order to maintain the links to its neighbour. Moreover, at the creation of the new blank cell, two new links are utilized: $u$ is the update channel of the new blank cell, and $l$ will later be used as the link to generate another cell. Thus a new cell is generated from $\overline{c}\langle t,l,r,u,\Box\rangle.\nil$, and the cell generator on the left is updated by $\overline{b_l}\langle l,t\rangle.\nil$. In the second case, a symmetrical procedure is implemented by $B_r(t,l,r,u)$.

Throughout the simulation of an RTM, the number of parallel components modelling individual tape cells will grow. We shall presuppose a numbering of these parallel components with consecutive integers from some interval $[m,n]$ ($m$ and $n$ are integers such that $m\leq n$), in agreement with the link structure. The numbering is reflected by a naming scheme that adds the subscript $i$ to the links $t$, $l$, $r$, $u$ and $d$ of the $i$th cell. We abbreviate $C(t_i,l_i,r_i,u_i,d_i)$ by $C_i(d_i)$, and $B_l(t_i,l_i,r_i,u_i)$ and $B_l(t_i,l_i,r_i,u_i)$ by $B_{l,i}$ and $B_{r,i}$, respectively. Let $\vec{d}_{[m,n]}=d_m,d_{m+1},\dots,d_{n-1},d_n$; we define:
\begin{equation*}
  \textit{Cells}_{[m,n]}(\vec{d}_{[m,n]}) \ddef (b_l, b_r, c)(B_{l,m-1} \parc{  C_m(d_m)} \parc{ C_{m+1}(d_{m+1})} \parc{ \cdots} \parc{ C_{n-1}(d_{n-1})} \parc{ C_n(d_n) }\parc{ B_{r,n+1}} \parc {! C} \parc {! B})
\enskip.
\end{equation*}

The component modelling the tape head serves as the interface between the tape cells and the RTM-specific control process.
It is defined as:
\begin{eqnarray*}
H &\ddef& h(t,l,r,u,d).H(t,l,r,u,d)\\
H(t,l,r,u,d)&\ddef&\overline{\textit{read}}\,d.\overline{h}\langle t,l,r,u,d\rangle.\nil+\textit{write}(e).\overline{u}\, e.\overline{h}\langle t,l,r,u,e\rangle.\nil\\
 &+& \textit{left}.l(l',r',u',d').\overline{h}\langle l,l',r',u',d'\rangle.\nil\\
 &+& \textit{right}.r(l',r',u',d').\overline{h}\langle r,l',r',u',d'\rangle.\nil
 \enskip.
\end{eqnarray*}

The tape head maintains two links to the current cell (a communication channel $t$ and an update channel $u$), as well as links to its left and right neighbour cells ($l$ and $r$, respectively). Furthermore, the tape head remembers the datum $d$ in the current cell. The datum $d$ may be output along the $\textit{read}$-channel. Furthermore, a new datum $e$ may be received  through the $\textit{write}$-channel, which is then forwarded through the update channel $u$ to the current cell. Finally, the tape head may receive instructions to move left or right, which has the effect of receiving information about the left or right neighbours of the current cell through $l$ or $r$, respectively. In all cases, a new incarnation of the tape head is started, with a call on the $h$-channel.

Let $\vec{t}_{[m,n]}=t_m,t_{m+1},\dots,t_{n-1},t_n$, let $\vec{u}_{[m,n]}=u_m,u_{m+1},\dots,u_{n-1},u_n$, and let $H_i=H(t_i,l_i,r_i,u_i,d_i)$; we define
\begin{equation*}
  \textit{Tape}_{[m,n]}^i(\vec{d}_{[m,n]}) \ddef (\vec{t}_{[m-1,n+1]}, \vec{u}_{[m,n]})((h)(H_i\parc {!H}) \parc{ \textit{Cells}_{[m,n]}(d_{[m,n]})})
\enskip.
\end{equation*}

\begin{lemma}\label{lemma:para-comp}
Suppose $C_i,B_{l,m},B_{r,n},H_i$ are as defined before, then the following statements are valid:
\begin{enumerate}
    \item $(c)(\overline{c}\langle t_i,l_i,r_i,u_i,d_i\rangle.\nil\parc{!C})\bbisimd (c)(C_i(d_i)\parc{!C})$
    \item $(b_l,b_r)(\overline{b_l}\langle t_m,r_m\rangle.\nil\parc{!B})\bbisimd (b_l,b_r,u_m,l_m)(B_{l,m}\parc{!B})$
    \item $(b_l,b_r)(\overline{b_r}\langle t_n,l_n\rangle.\nil\parc{!B})\bbisimd (b_l,b_r,u_n,r_n)(B_{r,n}\parc{!B})$
    \item $(h)(\overline{h}\langle t_i,l_i,r_i,u_i,d_i\rangle.\nil\parc{!H})\bbisimd (h)(H_i\parc{!H})$
\end{enumerate}
\end{lemma}
\fullversion{%
\begin{proof}
    We just show the first statement. There is only one $\tau$ transition from $(c)(\overline{c}\langle t_i,l_i,r_i,u_i,d_i\rangle.\nil\parc{!C})$, which exactly leads to $(c)(C_i(d_i)\parc{!C})$, which is state-preserving.
\end{proof}}

We shall write $P\step{a}\bbisimd P'$ for ``there is a $P''$ such that $P\step{a}P''$ and $P''\bbisimd P'$''.

\begin{lemma}\label{lemma:tape-behaviour}
There are four types of transitions from $\textit{Tape}_{[m,n]}^i(\vec{d}_{[m,n]})$:
\begin{enumerate}
    \item $\textit{Tape}_{[m,n]}^i(\vec{d}_{[m,n]})\step{\overline{\textit{read}}\,d_i}\bbisimd \textit{Tape}_{[m,n]}^i(\vec{d}_{[m,n]})$;
    \item $\textit{Tape}_{[m,n]}^i(\vec{d}_{[m,n]})\step{\textit{write}(e)}\bbisimd \textit{Tape}_{[m,n]}^i(d_{[m,i-1]},e,d_{[i+1,n]})$;
    \item $\textit{Tape}_{[m,n]}^i(\vec{d}_{[m,n]})\step{\textit{left}}\bbisimd \textit{Tape}_{[m,n]}^{i-1}(\vec{d}_{[m,n]})$ (if $i> m$);

    $\textit{Tape}_{[m,n]}^i(\vec{d}_{[m,n]})\step{\textit{left}}\bbisimd \textit{Tape}_{[m-1,n]}^{i-1}(\Box,\vec{d}_{[m,n]})$ (if $i= m$);
    \item $\textit{Tape}_{[m,n]}^i(\vec{d}_{[m,n]})\step{\textit{right}}\bbisimd \textit{Tape}_{[m,n]}^{i+1}(\vec{d}_{[m,n]})$ (if $i<n$);

    $\textit{Tape}_{[m,n]}^i(\vec{d}_{[m,n]})\step{\textit{right}}\bbisimd \textit{Tape}_{[m,n+1]}^{i+1}(\vec{d}_{[m,n]},\Box)$ (if $i= n$).
\end{enumerate}
\end{lemma}
\fullversion{%
\begin{proof}
 $\textit{Tape}_{[m,n]}^i(\vec{d}_{[m,n]})$ has four possible outgoing transitions (with labels $read$, $write$, $left$, $right$). Let them with $T'$ as the target.
We just argue (only in the first case) that $T'$ is indeed bisimilar to the correct process.
\begin{equation*}
    \textit{Tape}_{[m,n]}^i(\vec{d}_{[m,n]})\step{\overline{\textit{read}}\,d_i} (\vec{t}_{[m-1,n+1]}, \vec{u}_{[m,n]})((h)(\overline{h}\langle t_i,l_i,r_i,u_i,d_i\rangle\parc{!H}) \parc{ \textit{Cells}_{[m,n]}})=T'.
    \enskip.
\end{equation*}
         By Lemma~\ref{lemma:para-comp}, we have $(h)(\overline{h}\langle t_i,l_i,r_i,u_i,d_i\rangle\parc{!H})\bbisimd (h)(H_i\parc{!H})$, so by Lemma~\ref{lemma:pi-compat}, we have $T'\bbisimd \textit{Tape}_{[m,n]}^i(\vec{d}_{[m,n]})$.
\end{proof}}

\subsection{Finite control}\label{subsec:fincontrol}

We associate with every RTM $\M=(\Sta_{\M},\step{}_{\M},\uparrow_{\M})$ a finite specification of its control process. Here $m$ can be either $\textit{left}$ or $\textit{right}$.
\begin{eqnarray}
S&\ddef& \sum_{s\in \Sta_{\M}}s.\sum_{d\in\D_{\Box}}d.S_{s,d}\nonumber\\
S_{s,d}&\ddef&\sum_{(s,d,a,e,m,t)\in\step{}_{\M}}a.\overline{\textit{write}}\,e.\overline{m}.\textit{read}(f).\overline{t}.\overline{f}.\nil\nonumber
\end{eqnarray}


Let $\vec{s}=s_1,s_2,\ldots,s_m\in \Sta_{\M}$, and $\vec{e}=e_1,e_2,\ldots,e_n\in \D_{\Box}$; we define
\begin{equation*}
  \textit{Control}_{s,d} \ddef (\vec{s}, \vec{e})(S_{s,d}\parc{!S})
\enskip.
\end{equation*}

 The following lemma illustrates the behaviour of the control process.

\begin{lemma}~\label{lemma:transition-step}
Given an RTM $\M=(\Sta_{\M},\step{}_{\M},\uparrow_{\M})$, we have the following transition sequence:
 \begin{equation*}
\textit{Control}_{s,d}\step{a}(\vec{s}, \vec{e})(\overline{\textit{write}}\,e.\overline{m}.\textit{read}(f).\overline{t}.\overline{f}.\nil\parc{!S})\step{\overline{\textit{write}}\,e}\step{\overline{m}}\step{\textit{read}\,f} \bbisimd\textit{Control}_{t,f}.\nil
\enskip.
\end{equation*}
if and only if there is a transition rule $(s,d,a,e,m,t)\in\step{}_{\M}$.
\end{lemma}

Finally, for a given RTM $\M$, we associate with every configuration $(s,\delta_L\check{d}\delta_R)$ a $\pi$-term $M_{s,\delta_L\check{d}\delta_R}$, consisting of a parallel composition of the specifications of its tape instance and control process. Let $\vec{r}=\textit{read},\textit{write},\textit{left},\textit{right}$; we define
 \begin{equation*}
  M_{s,\delta_L\check{d}\delta_R}=(\vec{r})(\textit{Control}_{s,d}\parc{ \textit{Tape}^{i}_{[m,n]}(\vec{d}_{[m,n]})}),\,where\, \vec{d}_{[m,n]}=\delta_L\check{d}\delta_R
\enskip.
\end{equation*}

 The following lemma shows that $M_{s,\delta_L\check{d}\delta_R}$ actually simulates every computation step of an RTM.

 \begin{lemma}\label{lemma:behaviour-m}
 Given an RTM $\M=(\Sta_{\M},\step{}_{\M},\uparrow_{\M})$, for every configuration $(s,\delta_L\check{d}\delta_R)$, its specification $M_{s,\delta_L\check{d}\delta_R}$ has the following transition
   \begin{equation*}
   M_{s,\delta_L\check{d}\delta_R}\step{a}\bbisimd M_{t,\delta_L'\check{f}\delta_R'}
   \enskip,
   \end{equation*}
   if and only if there is a transition $(s,\delta_L\check{d}\delta_R)\step{a}(t,\delta_L'\check{f}\delta_R')$.
 \end{lemma}
\fullversion{%
 \begin{proof}
 Suppose the transition is resulted from the rule $(s,d,a,e,m,t)\in\step{}_{\M}$, then according to Lemma~\ref{lemma:transition-step}, we have
\begin{equation*}
M_{s,\delta_L\check{d}\delta_R}\step{a}(\vec{r},\vec{s},\vec{e})(\overline{\textit{write}}\,e.\overline{m}.\textit{read}(f).\overline{t}.\overline{f}.\nil\parc{!S}\parc{\textit{Tape}^{i}_{[m,n]}(\vec{d}_{[m,n]})})=M'
\enskip.
\end{equation*}

 Then we just prove that
 \begin{equation*}
 M'\bbisimd (\vec{r},\vec{s},\vec{e})(\overline{m}.\textit{read}(f).\overline{t}.\overline{f}.\nil\parc{!S}\parc{\textit{Tape}^{i}_{[m,n]}(d_m,\ldots,d_{i-1}, e,d_{i+1},\ldots,d_n)})=M''
 \enskip.
  \end{equation*}
  Applying Lemma~\ref{lemma:tape-behaviour}, we get
  \begin{equation*}
  M'\step{\tau} (\vec{r},\vec{s},\vec{e})(\overline{m}.\textit{read}(f).\overline{t}.\overline{f}.\nil\parc{!S}\parc{T'})
   \enskip,
  \end{equation*} where $T'\bbisimd \textit{Tape}^{i}_{[m,n]}(d_m,\ldots,d_{i-1}, e,d_{i+1},\ldots,d_n)$.
   Thus $M'\bbisimd M''$ according to Lemma~\ref{lemma:pi-compat}.

 Hence, let $T''\bbisimd \textit{Tape}^{i'}_{[m',n']}(\vec{d'}_{[m',n']}),\,where\, \vec{d'}_{[m',n']}=\delta'_L\check{f}\delta'_R$, and we get a transition sequence,
 \begin{eqnarray*}
 M'&\step{\tau}\bbisimd&(\vec{r},\vec{s},\vec{e})(\overline{m}.\textit{read}(f).\overline{t}.\overline{f}.\nil\parc{!S}\parc{T'})\\
 &\step{\tau}\bbisimd&(\vec{r},\vec{s},\vec{e})(\textit{read}(f).\overline{t}.\overline{f}.\nil\parc{!S}\parc{T''})\\
 &\step{\tau}\bbisimd&(\vec{r},\vec{s},\vec{e})(\overline{t}.\overline{f}.\nil\parc{!S}\parc{T''})\\
 &\step{}^{*}\bbisimd& M_{t,\delta_L'\check{f}\delta_R'}
 \end{eqnarray*}
   by applying Lemma~\ref{lemma:tape-behaviour}.
 \end{proof}}

\begin{theorem}\label{thm:rtm-pi-spec}
Given an RTM $\M$, we have
\begin{equation*}
\T(M_{\uparrow,\check{\Box}})\bbisimd\T(\M)
\enskip.
\end{equation*}
\end{theorem}
\fullversion{%
\begin{proof}
Let $\M=(\Sta,\step{}',\uparrow)$, we replace every $\tau$-transition to an $i$-labeled transition, where $i\notin\Atau$ indicates an inner step from the RTM rules, and get an RTM $\M'=(\Sta,\step{}',\uparrow)$. Thus we can establish a specification $M_{\uparrow,\check{\Box}}'$, for the initial state $(\uparrow,\check{\Box})$ of $\M'$, we have

We establish a relation
\begin{equation*}
\R'=
\{(M_{s,\delta_L\check{d}\delta_R}',(s,\delta_L\check{d}\delta_R))\mid s\in \Sta,\,\delta_L,\,\delta_R\in \Dbox^{*},\,\check{d}\in\check{\Dbox}\}
\enskip.
\end{equation*}
Using Lemma~\ref{lemma:behaviour-m}, it enough to show that $\R'$ is a branching bisimulation up to $\bbisim$. Thus we get $\T(M_{\uparrow,\check{\Box}}')\bbisim\T((\uparrow,\check{\Box}))$ by Lemma~\ref{lemma:up-to}.

Now we proceed to show this branching bisimilarity is divergence preserving.

Note that there is no $\tau$-transition in $\M'$, which means $\T((\uparrow,\check{\Box}))$ has no divergence. Then, by Lemma~\ref{lemma:behaviour-m}, the specification of a certain configuration $M_{s,\delta_L\check{d}\delta_R}'$ can only do some $a$-labelled transitions, where $a\in\A\cup \{i\}$, i.e.
\begin{equation*}
   M_{s,\delta_L\check{d}\delta_R}'\step{a}M'\bbisimd M_{t,\delta_L'\check{f}\delta_R'}'
   \enskip.
\end{equation*}
Since there is no $\tau$ transition from the term $M_{t,\delta_L'\check{f}\delta_R'}'$, it follows that $M'$ has no divergence either. Hence, all the reachable terms from $M_{s,\delta_L\check{d}\delta_R}'$ introduce no divergence, and we have,

\begin{equation*}
\T(M_{\uparrow,\check{\Box}}')\bbisimd\T((\uparrow,\check{\Box}))
\enskip.
\end{equation*}

Finally, we switch back to $\M$, by changing all the $i$ labelled transition to $\tau$, and we let $M_{\uparrow,\check{\Box}}$ be the specification of the initial state of $\M$. We can also establish that the relation

\begin{equation*}
\R=
\{(M_{s,\delta_L\check{d}\delta_R},(s,\delta_L\check{d}\delta_R))\mid s\in \Sta,\,\delta_L,\,\delta_R\in \Dbox^{*},\,\check{d}\in\check{\Dbox}\}
\enskip.
\end{equation*}

is a branching bisimulation up to $\bbisim$. Moreover, note that every infinite sequence of the form $\step{i}\step{}^{*}\step{i}\step{}^{*}\cdots$ from $M_{\uparrow,\check{\Box}}'$ corresponds with a infinite sequence of the form $\step{i}\step{i}\cdots$ from $\M'$, and vice versa. Additionally, there is no divergence from $M_{\uparrow,\check{\Box}}'$. Therefore, every infinite $\tau$ labelled sequence from $M_{\uparrow,\check{\Box}}$ corresponds with a infinite $\tau$ labelled sequence from $\M$. So we can conclude that $\R\subseteq\bbisimd$.
\end{proof}}


Thus we have the following expressiveness result for the $\pi$-calculus.
\begin{corollary}\label{coro:pi-exp}
For every executable transition system $T$ there exists a $\pi$-term $P$, such that $T\bbisimd\T(P)$.
\end{corollary}

\section{Executability of the $\pi$-Calculus}\label{sec:exe}

We have proved that every executable behaviour can be specified in the $\pi$-calculus modulo divergence-preserving branching bisimilarity. We shall now investigate to what extent behaviour specified in the $\pi$-calculus is executable. Recall that we have defined executable behaviour as behaviour of an RTM. So, in order to prove that the behaviour specified by a $\pi$-term is executable, we need to show that the transition system associated with this $\pi$-term is behaviourally equivalent to the transition system associated with some RTM.

 Note that there is an apparent mismatch between the formalisms of RTMs and the $\pi$-calculus. On the one hand, the notion of  RTM as we have defined in Section~\ref{sec:def} presupposes \emph{finite} sets $\Atau$ and $\Dbox$ of actions and data symbols, and also the transition relation of an RTM is \emph{finite}. As a consequence, we have observed, the transition system associated with an RTM is finitely branching, and, in fact, its branching degree is bounded by a natural number. (Note that this does not mean that RTMs cannot deal with data of unbounded size; it only means that it has to be encoded using finitely many symbols.) The $\pi$-calculus, on the other hand, presupposes an infinite set of names by which an infinite set of actions $\Api$ is generated. Furthermore, the transition system associated with a $\pi$-term by the structural operational semantics (see Fig.~\ref{tab:pi-semantics}) may contain states with an infinite branching degree, due to the rules for input prefix and bound output prefix. Regarding this gap, we shall explore two ways to establish simulation of $\pi$-calculus terms by RTMs. One is to extend the formalism of RTMs to presuppose an infinite set of actions, and the other is to restrict the $\pi$-calculus to use a finite set of names.

\subsection{RTMs with Infinitely Many Actions}
Let us first consider allowing RTMs to have infinitely many actions in order to accommodate for the infinitely many names in the $\pi$-calculus.

Recall Definition~\ref{def:rtm}, an RTM has a finite set of states $\Sta$ and a finite set of transition rules defining the associated transition relation. If we allow RTMs to have infinitely many actions, then, inevitably, we should also allow them to have infinitely many transition rules. The following lemma shows that we then also either need infinitely many states or infinitely many data symbols.

\begin{lemma}~\label{lemma:RTM-infi}
There does not exist an RTM with infinitely many actions but finitely many states and data symbols that simulates the $\pi$-term $P=x(y).\bar{y}.\nil$ modulo branching bisimilarity.
\end{lemma}
\fullversion{%
\begin{proof}
Suppose $\M=(\Sta,\step{},\uparrow)$ is an RTM such that $\T(\M)\bbisim\T(P)$.

The input transitions $P\step{xy_1}\overline{y_1}.\nil,\,P\step{xy_2}\bar{y_2}.\nil,\ldots$ lead to infinitely many states $\bar{y_1}.\nil,\bar{y_2}.\nil,\ldots$, which are all mutually distinct modulo branching bisimilarity.

  Let $C=(\uparrow,\check{\Box})$ be the initial configuration of $\M$. We have $C\bbisim P$, so $C$ admits the following transition sequences: $C\step{}^{*}\step{xy_{1}}\step{}^{*}C_1\step{\overline{y_{1}}},\,C\step{}^{*}\step{xy_{2}}\step{}^{*}C_2\step{\overline{y_{2}}},\ldots$, where $C_1\bbisim\bar{y_{1}}.\nil,\,C_2\bbisim\bar{y_{2}}.\nil,\ldots$.

The rules of an RTM are of the form $s\step{a[d/e]M}t$, where $s,\,t\in\Sta$, and $d,\,e\in\Dbox$; we call the pair $(s,d)$ the trigger of this rule. A configuration $(s',\delta_L\check{d'}\delta_R)$ satisfies the trigger $(s,d)$ if $s=s'$ and $d=d'$. Now observe that a rule $s\step{a[d/e]M}s'$ gives
rise to an $a$-transition from every configuration satisfying its trigger $(s,d)$. Since $\Sta$ and $\Dbox$ are finite sets, there are finitely many triggers.

So, in the infinite list of configurations $C_1,C_2,\ldots$, there are at least two configurations $C_i$ and $C_j$, satisfying the same trigger $(s,d)$; these configurations must have the same outgoing transitions.

Now we conclude that we cannot have $C_i\bbisim \bar{y_{i}}.\nil$ and $C_j\bbisim\bar{y_{j}}.\nil$. Since $C_j\step{y_j}$, we also have the transition $C_i\step{\bar{y_{j}}}$ triggered by $(s,d)$. Hence $C_i\not\bbisim  \bar{y_{i}}.\nil$, and we get a contradiction to $\T(\M)\bbisim\T(P)$.
\end{proof}}

Now, assume we allow the alphabet of data symbols to be infinite. It is then straightforward to use it to encode an infinite set of control states. Allowing an infinite set of data symbols, in fact, greatly enhances the expressiveness of RTMs, as the following theorem shows.

\begin{theorem}~\label{thm:RTM-infi}
Every infinitely branching effective transition system can be simulated up to divergence-preserving branching bisimilarity by an RTM with infinite sets of action symbols and data symbols.
\end{theorem}
\fullversion{%
\begin{proof}
 Let $T=(\Sta_T,\step{}_T,\uparrow_T)$ be an $\A$-labelled effective transition system, and let $\phi: \Sta_T\rightarrow \mathbb{N}$ be an injective function encoding its states as natural numbers. Then, an RTM with infinite sets of action symbols and data symbols $\M(T)=(\Sta,\step{},\uparrow)$ is defined as follows.
\begin{enumerate}
    \item $\Sta=\{s,t,\uparrow\}$ is the set of control states.
    \item $\step{}$ is a finite set of $(\Dbox\times\A\times\Dbox\times\{L,R\})$-labelled \emph{transition rules}, and it consists of the following transition rules:
        \begin{enumerate}
            \item $\uparrow\step{\tau[\Box/\phi(\uparrow_T)]R}s$
            \item $s\step{\tau[\Box/\Box]L}t$
            \item $t\step{a[\phi(s_1)/\phi(s_2)]R}s$ if there is a transition $s_1\step{a} s_2$ for states $s_1, s_2\in\Sta_T$.
        \end{enumerate}
    \item $\uparrow\in\Sta$ is a distinguished \emph{initial state}.
\end{enumerate}

Then the transition system $\T(\M(T))$ is divergence-preserving branching bisimilar to $T$.
Note that a transition step $s_1\step{a}s_2$ of the $\pi$-calculus is simulated by a sequence $(s,\phi(s_1)\check{\Box})\step{\tau}(t,\check{\phi(s_1)}\Box)\step{a}(s,\phi(s_2)\check{\Box})$
\end{proof}}

As a consequence, we can simulate every $\pi$-calculus term up to divergence-preserving branching bisimilarity with an RTM having infinite sets of action symbols and data symbols.
So, if we would extend the formalism of RTMs allowing infinitely many action symbols and
data symbols and define the notion of executability on the basis of it, then we would get that every $\pi$-calculus
process is executable up to divergence-preserving branching bisimilarity. One may argue, however, that such extension is not in accordance with reality, referring to the finiteness of realistic computing systems. Actually, this result only shows the existence of such theoretical models, rather than giving a way of implementation. Moreover, the conclusion is valid for every model with an effective operational semantics, even if it has infinite branching.

\subsection{Restricting the $\pi$-calculus}

Now we proceed to consider the other option, which is to propose a restriction on the transition systems associated with $\pi$-terms such that they refer only to finitely many actions.

The infinity of the set of actions in the $\pi$-calculus arises in two ways, the free input names and the bound output names. The free input names allow a process to receive any potential input from the environment and the bound output names give a process the ability to generate unboundedly many distinct private channels to communicate with other processes. For both purposes, infinite branching of the transition system is essential.
Observe, that the infinite branching caused by input prefix can be thought of as a technical device in the operational semantics to model the communication of an arbitrary name from one parallel component to another. The name that will be received, can either be a free name of the sending process (a value), or a restricted name (a private channel). Since the sending parallel component will only have a finite number of free names, only finitely many values can be communicated. Although, technically speaking, according to the operational semantics, infinitely many distinct private channels may be communicated when an input prefix synchronises with a bound output prefix, the communicated private channel is not observable, and the resulting $\pi$-terms are all equated by $\alpha$-conversion, so the only observable effect of the interaction is that after the communication the sending and receiving parties share a private channel of which the name is irrelevant.

Our goal is to investigate to what extent the behaviour specified by an individual $\pi$-term is executable. Motivated by the above intuitive interpretation of interaction of a $\pi$-term with its environment, we assume that the behaviour specified by that $\pi$-term is executed in an environment that may offer data values from some presupposed finite set on its input channels. This assumption seems reasonable as a machine should know in advance which symbols to expect as an input. Furthermore, we assume that there is a facility for establishing a private channel between the $\pi$-term and its environment. (Such a facility could, e.g., be implemented using encryption, but we will abstract from the actual implementation of the facility.) We define a restriction on the transition systems associated with $\pi$-terms that is based on these assumptions.

\begin{definition}\label{def:lts-pi-restrict}
Let $\N'\subseteq \N$ be a set of names, let $\Api'=\Api - (\{xy\mid x,y\in\N,\ y\notin\N'\}\cup\{\overline{x}(z)\mid x, z\in\N\})$, and let $P$ be a $\pi$-term. The transition system associated with $P$ restricted to $\N'$ , denoted by $\T(P)\upharpoonright \N'$, is a triple $(\Sta_P\upharpoonright \N',\step{}_{P}\upharpoonright \N',P)$, obtained from $\T(P)=(\Sta_P,\step{}_P,P)$ as follows:
\begin{enumerate}
    \item $\Sta_P\upharpoonright \N'$ is the set of states reachable from $P$ by means of transitions that are not labelled by $xy$ ($y\notin\N'$); and
    \item $\step{}_{P}\upharpoonright \N'$ is the restriction of $\step{}_P$ obtained by excluding all transitions labelled with $xy$ ($y\notin\N')$, and relabelling all transitions labelled with $\overline{x}(z)$ ($x,z\in\N$) to $\mathbf{\nu}\overline{x}$, i.e.,
        \begin{multline*}
           \step{}_P\upharpoonright \N'=({\step{}_P} \cap (\Sta_P\upharpoonright \N'\times\Api'\times\Sta_P\upharpoonright \N'))\cup\{(s,\mathbf{\nu}\overline{x},t)\mid s,t\in\Sta_P\upharpoonright \N', s\step{\overline{x}(z)}_P t\}
        \enskip.
        \end{multline*}
\end{enumerate}
\end{definition}

Using \cite[Lemma 1.4.1]{SW01}, it is straightforward to show that for every $\pi$-term the set of actions of the $\pi$-calculus appearing as labels in $\T(P)\upharpoonright \N'$ is finite. Furthermore, the transition system associated with a $\pi$-term by the operational semantics, and also its restriction according to Definition~\ref{def:lts-pi-restrict} are clearly effective. Hence, as an immediate corollary of Theorem~\ref{thm-blt}(3), we may conclude that the transition system associated  with a $\pi$-term can be simulated by an RTM  modulo (divergence-insensitive) branching bisimilarity.

\begin{corollary}\label{coro:pi-exe}
  For every closed $\pi$-term $P$, and for every finite set of input names $\N'\subseteq\N$, there exists an RTM $M$ such that $\T(P)\upharpoonright \N'\bbisim\T(M)$.
\end{corollary}

The following example shows that there exist $\pi$-terms with which the structural operational semantics associates a transition system without divergence that is unboundedly branching up to $\bbisimd$. Note that by Theorem~\ref{thm:non-exe} such $\pi$-terms are not executable modulo divergence-preserving branching bisimilarity.

\begin{example}\label{exp:pi-unbound}
Consider the $\pi$-process $P\ddef (c,i,d,s,\mathit{flip})({\overline{i}\,s.\nil} \parc {\mathit{flip}.\nil} \parc {!C} \parc {!I} \parc {!D})$, with $C$, $I$ and $D$ defined as follows:
\begin{eqnarray*}
C&\ddef& c(h,t,b).(\overline{h}\langle t,b\rangle.\nil+\mathit{flip}.\overline{c}\langle h,t,1\rangle.\nil)\\
I&\ddef& i(h).(\mathit{inc}.(h')\overline{c}\langle h',h,0 \rangle.\overline{i}\, h'.\nil + \mathit{flush}.\overline{\mathit{flip}}.\overline{d}\, h.\nil)\\
D&\ddef& d(h).(h(t,b).\overline{b}.\overline{d}\, t.\nil)
\end{eqnarray*}

Intuitively, the process $!C$ facilitates the generation of a linked list of one-bit cells with a pointer $h$ to the head of the list, a pointer $t$ to the tail of the list, and a bit $b$. Each cell may either output, along $h$, the link $t$ to the tail of the list and its bit $b$, or it may receive the instruction $\mathit{flip}$ after which it recreates itself with the value $1$. The process $I$ serves as the interface process. It maintains a link to the head of the list. Upon receiving an $\mathit{inc}$-instruction, it adds another one-bit cell to the list, and upon receiving the $\mathit{flush}$-instruction, it flips at most one of the bits, and then calls $D$. The process $D$ then simply outputs the bits in reverse sequence.

Consider the state reached after performing $n$ $\mathit{inc}$-actions, followed by a $\mathit{flush}$-action. In this state, the list contains a string of $n$ $0$s. The $\tau$-transitions that correspond to the interaction of $\overline{\mathit{flip}}$ between $I$ and one of the $\mathit{flip}$s of one of the one-bit cells or $\mathit{flip}$ in the definition of $P$ have the effect of non-deterministically changing (at most) one of the $0$s to a $1$. Note that there are $n+1$ such $\tau$-transitions, and since $D$ will subsequently output the sequence in order, the states reached by these $\tau$-transitions are (pairwise) not divergence-preserving branching bisimilar. Hence, it follows that for every $n$, the transition system associated with $P$ has a reachable state with a branching degree modulo $\bbisimd$ of at least $n+1$. It follows that the transition system associated with $P$ is unboundedly branching up to $\bbisimd$.

\end{example}

Note that the only names occurring as part of the labels on the transitions in the transition system associated with the $\pi$-term $P$ in the preceding example are $\overline{0}$, $\overline{1}$, $\mathit{inc}$ and $\mathit{flush}$, so if $\N'$ contains at least these four names, then $P$ satisfies $\T(P)\upharpoonright\N'=\T(P)$. Let us say, in general, that a $\pi$-term $P$ has \emph{finitely many observable names} if there exists a finite set $\N'\subseteq\N$ such that $\T(P)\upharpoonright\N'=\T(P)$. Note that, in this case, $P$ cannot have parameterised free inputs, nor bound outputs. For $\pi$-terms with finitely many observable names, we have the following corollary as a consequence of a combination of Corollary~\ref{coro:pi-exe} and Example~\ref{exp:pi-unbound}.

\begin{corollary}\label{cor:pi-non-exe}
  Every closed $\pi$-term $P$ with finitely many observable names is executable up to (divergence-insensitive) branching bisimilarity, but there exist closed $\pi$-terms with finitely many observable names that are not executable up to divergence-preserving branching bisimilarity.
\end{corollary}

Our notion of restriction is introduced to restrict labelled transition systems associated with $\pi$-terms to finitely many names. Alternatively, we could define a finite version $\pi_{\textit{fin}}$ of the $\pi$-calculus, presupposing a \emph{finite} set of names $\N$ right from the beginning. If $\T(P)$ is the labelled transition system associated with $P$ according to the operational semantics of $\pi_{\textit{fin}}$, then $\T(P)\upharpoonright \N$ is obtained from $\T(P)$ by replacing all transitions with the label $\overline{x}(z)$ by transitions with the label $\nu\overline{x}$. Apart from this modification, restriction keeps all observable behaviour.

\section{Conclusions and Related Work}\label{sec:conclusions}


We have investigated the expressiveness of the $\pi$-calculus in relation to the theory of executability provided by reactive Turing machines. The issue of the expressiveness of the $\pi$-calculus has been extensively studied (see \cite{Gorla10} for a comprehensive overview of research in this area). A distinction is usually made between absolute and relative expressiveness results. The absolute expressiveness results focus on proving the (im)possibility of expressing a computational phenomenon in a calculus; the relative expressiveness results are mostly about encoding one calculus in another. Our results pertain to the absolute expressiveness of the $\pi$-calculus.

We have established that, up to divergence-preserving branching bisimilarity, every executable transition system can be specified in the $\pi$-calculus, showing that the $\pi$-calculus is reactively Turing powerful. Milner already established in~\cite{Milner1990} that the $\pi$-calculus is Turing powerful, by exhibiting an encoding of the $\lambda$-calculus in the $\pi$-calculus by which every reduction in the $\lambda$-calculus is simulated by a sequence of reductions in the $\pi$-calculus. Our result that all executable behaviour can be specified in the $\pi$-calculus up to divergence-preserving branching bisimilarity also implies that the $\pi$-calculus is Turing powerful, and thus it subsumes Milner's result. Similarly, in \cite{Busi2009} several expressiveness results for variants of CCS are obtained via an encoding of Random Access Machines, and also those results only make claims about the computational expressiveness of the calculi. Notice that the results in \cite{Milner1990} and \cite{Busi2009} confirm the computational power of the respective calculi, but do not make a qualitative statement about its interactive expressiveness. By showing that reactive Turing machines can be faithfully simulated, we at the same time confirm the interactive expressiveness of the $\pi$-calculus.

In his recent work~\cite{Fu14}, Fu also proposes to study computation and interaction in an integrated theory. His theory is built on four fundamental principles, rather than on a machine model. One of the contributions of his theory is a calculus including a bare minimum of primitives to be computationally and interactively complete, and he uses it to confirm the completeness of the $\pi$-calculus. We leave it for future work to explore the relationship between Fu's theory of interaction and the theory of executability based on reactive Turing machines.

We have observed that it is possible to specify behaviour in the $\pi$-calculus that is not executable up to any reasonable notion of behavioural equivalence, simply because it uses infinitely many observable names. For the presentation of the $\pi$-calculus it is technically important to presuppose an infinite set of names especially to model the feature of dynamic creation of private channels between components. Allowing RTMs to have an infinite set of actions and either an infinite set of states or an infinite set of data symbols would arguably lead to an unrealistically powerful notion of executability. Moreover, in a real system, private channels between components are likely to be implemented differently, e.g., using some form of encryption. An interesting idea for future research is to consider a notion of RTM with atoms along the lines of \cite{BKLT13}, which possibly leads to a more realistic theory of executability allowing infinitely many actions. In this paper, we have shown that a behaviour specified in the $\pi$-calculus is executable up to the divergence-insensitive variant of branching bisimilarity if one restricts to finitely many observable names and does not associate a unique identifier with every dynamically created private channel.

It has been claimed (e.g., in \cite{eberbach2007}) that the $\pi$-calculus provides a model of computation that is behaviourally more expressive than Turing machines. Our results provide further justification for this claim, and characterise the difference. It should be noted that the difference in expressive power is at the level of interaction (allowing interaction between an unbounded number of components), rather than at the level of computation. 

\paragraph{Acknowledgement} We thank the reviewers for the elaborate discussions on the ICE 2015 forum, which led to several improvements of the article.


\begin{thebibliography}{10}
\providecommand{\bibitemdeclare}[2]{}
\providecommand{\surnamestart}{}
\providecommand{\surnameend}{}
\providecommand{\urlprefix}{Available at }
\providecommand{\url}[1]{\texttt{#1}}
\providecommand{\href}[2]{\texttt{#2}}
\providecommand{\urlalt}[2]{\href{#1}{#2}}
\providecommand{\doi}[1]{doi:\urlalt{http://dx.doi.org/#1}{#1}}
\providecommand{\bibinfo}[2]{#2}

\bibitemdeclare{article}{Baeten1987129}
\bibitem{Baeten1987129}
\bibinfo{author}{J.~C.~M. \surnamestart Baeten\surnameend},
  \bibinfo{author}{J.~A. \surnamestart Bergstra\surnameend} \&
  \bibinfo{author}{J.~W. \surnamestart Klop\surnameend} (\bibinfo{year}{1987}):
  \emph{\bibinfo{title}{{On the consistency of Koomen's Fair Abstraction
  Rule}}}.
\newblock {\sl \bibinfo{journal}{Theoretical Computer Science}}
  \bibinfo{volume}{51}(\bibinfo{number}{1–2}), pp. \bibinfo{pages}{129 --
  176}, \doi{10.1016/0304-3975(87)90052-1}.

\bibitemdeclare{book}{baeten2010}
\bibitem{baeten2010}
\bibinfo{author}{Jos C.~M. \surnamestart Baeten\surnameend},
  \bibinfo{author}{Twan \surnamestart Basten\surnameend} \&
  \bibinfo{author}{Michel~A. \surnamestart Reniers\surnameend}
  (\bibinfo{year}{2010}): \emph{\bibinfo{title}{Process algebra: equational
  theories of communicating processes}}.
\newblock \bibinfo{volume}{50}, \bibinfo{publisher}{Cambridge University
  Press}, \doi{10.1017/CBO9781139195003}.

\bibitemdeclare{article}{BLT2013}
\bibitem{BLT2013}
\bibinfo{author}{Jos C.~M. \surnamestart Baeten\surnameend},
  \bibinfo{author}{Bas \surnamestart Luttik\surnameend} \&
  \bibinfo{author}{Paul \surnamestart van Tilburg\surnameend}
  (\bibinfo{year}{2013}): \emph{\bibinfo{title}{{Reactive Turing Machines}}}.
\newblock {\sl \bibinfo{journal}{Inform. Comput.}} \bibinfo{volume}{231}, pp.
  \bibinfo{pages}{143--166}, \doi{10.1016/j.ic.2013.08.010}.

\bibitemdeclare{article}{Bas96}
\bibitem{Bas96}
\bibinfo{author}{Twan \surnamestart Basten\surnameend} (\bibinfo{year}{1996}):
  \emph{\bibinfo{title}{Branching Bisimilarity is an Equivalence Indeed!}}
\newblock {\sl \bibinfo{journal}{Inf. Process. Lett.}}
  \bibinfo{volume}{58}(\bibinfo{number}{3}), pp. \bibinfo{pages}{141--147},
  \doi{10.1016/0020-0190(96)00034-8}.

\bibitemdeclare{inproceedings}{BKLT13}
\bibitem{BKLT13}
\bibinfo{author}{Mikolaj \surnamestart Bojanczyk\surnameend},
  \bibinfo{author}{Bartek \surnamestart Klin\surnameend},
  \bibinfo{author}{Slawomir \surnamestart Lasota\surnameend} \&
  \bibinfo{author}{Szymon \surnamestart Torunczyk\surnameend}
  (\bibinfo{year}{2013}): \emph{\bibinfo{title}{Turing Machines with Atoms}}.
\newblock In: {\sl \bibinfo{booktitle}{28th Annual {ACM/IEEE} Symposium on
  Logic in Computer Science, {LICS} 2013, New Orleans, LA, USA, June 25-28,
  2013}}, \bibinfo{publisher}{{IEEE} Computer Society}, pp.
  \bibinfo{pages}{183--192}, \doi{10.1109/LICS.2013.24}.

\bibitemdeclare{article}{Busi2009}
\bibitem{Busi2009}
\bibinfo{author}{Nadia \surnamestart Busi\surnameend},
  \bibinfo{author}{Maurizio \surnamestart Gabbrielli\surnameend} \&
  \bibinfo{author}{Gianluigi \surnamestart Zavattaro\surnameend}
  (\bibinfo{year}{2009}): \emph{\bibinfo{title}{On the expressive power of
  recursion, replication and iteration in process calculi}}.
\newblock {\sl \bibinfo{journal}{Mathematical Structures in Computer Science}}
  \bibinfo{volume}{19}(\bibinfo{number}{06}), pp. \bibinfo{pages}{1191--1222},
  \doi{10.1017/S096012950999017X}.

\bibitemdeclare{article}{eberbach2007}
\bibitem{eberbach2007}
\bibinfo{author}{Eugene \surnamestart Eberbach\surnameend}
  (\bibinfo{year}{2007}): \emph{\bibinfo{title}{The \$-calculus process algebra
  for problem solving: A paradigmatic shift in handling hard computational
  problems}}.
\newblock {\sl \bibinfo{journal}{Theoretical Computer Science}}
  \bibinfo{volume}{383}(\bibinfo{number}{2}), pp. \bibinfo{pages}{200--243},
  \doi{10.1016/j.tcs.2007.04.012}.

\bibitemdeclare{unpublished}{Fu14}
\bibitem{Fu14}
\bibinfo{author}{Yuxi \surnamestart Fu\surnameend} (\bibinfo{year}{2014}):
  \emph{\bibinfo{title}{Theory of Interaction}}.
\newblock \urlprefix\url{http://basics.sjtu.edu.cn/~yuxi/}.

\bibitemdeclare{article}{Fu2010}
\bibitem{Fu2010}
\bibinfo{author}{Yuxi \surnamestart Fu\surnameend} \& \bibinfo{author}{Hao
  \surnamestart Lu\surnameend} (\bibinfo{year}{2010}): \emph{\bibinfo{title}{On
  the expressiveness of interaction}}.
\newblock {\sl \bibinfo{journal}{Theoretical Computer Science}}
  \bibinfo{volume}{411}(\bibinfo{number}{11–13}), pp. \bibinfo{pages}{1387 --
  1451}, \doi{10.1016/j.tcs.2009.11.011}.

\bibitemdeclare{article}{Glabbeek1996}
\bibitem{Glabbeek1996}
\bibinfo{author}{R.~J. \surnamestart van Glabbeek\surnameend} \&
  \bibinfo{author}{W.~Peter \surnamestart Weijland\surnameend}
  (\bibinfo{year}{1996}): \emph{\bibinfo{title}{Branching time and abstraction
  in bisimulation semantics}}.
\newblock {\sl \bibinfo{journal}{Journal of the ACM (JACM)}}
  \bibinfo{volume}{43}(\bibinfo{number}{3}), pp. \bibinfo{pages}{555--600},
  \doi{10.1145/233551.233556}.

\bibitemdeclare{inproceedings}{Glabbeek1993}
\bibitem{Glabbeek1993}
\bibinfo{author}{Rob~J. \surnamestart van Glabbeek\surnameend}
  (\bibinfo{year}{1993}): \emph{\bibinfo{title}{{The linear time — branching
  time spectrum II}}}.
\newblock In: {\sl \bibinfo{booktitle}{CONCUR'93}},
  \bibinfo{organization}{Springer}, pp. \bibinfo{pages}{66--81},
  \doi{10.1007/3-540-57208-2\_6}.

\bibitemdeclare{article}{Glabbeek2009}
\bibitem{Glabbeek2009}
\bibinfo{author}{Rob~J. \surnamestart van Glabbeek\surnameend},
  \bibinfo{author}{Bas \surnamestart Luttik\surnameend} \&
  \bibinfo{author}{Nikola \surnamestart Tr{\v{c}}ka\surnameend}
  (\bibinfo{year}{2009}): \emph{\bibinfo{title}{Branching bisimilarity with
  explicit divergence}}.
\newblock {\sl \bibinfo{journal}{Fundamenta Informaticae}}
  \bibinfo{volume}{93}(\bibinfo{number}{4}), pp. \bibinfo{pages}{371--392},
  \doi{10.3233/FI-2009-109}.

\bibitemdeclare{article}{Gorla10}
\bibitem{Gorla10}
\bibinfo{author}{Daniele \surnamestart Gorla\surnameend}
  (\bibinfo{year}{2010}): \emph{\bibinfo{title}{Towards a unified approach to
  encodability and separation results for process calculi}}.
\newblock {\sl \bibinfo{journal}{Inf. Comput.}}
  \bibinfo{volume}{208}(\bibinfo{number}{9}), pp. \bibinfo{pages}{1031--1053},
  \doi{10.1016/j.ic.2010.05.002}.

\bibitemdeclare{incollection}{Milner1990}
\bibitem{Milner1990}
\bibinfo{author}{Robin \surnamestart Milner\surnameend} (\bibinfo{year}{1990}):
  \emph{\bibinfo{title}{Functions as processes}}.
\newblock In \bibinfo{editor}{Michael~S. \surnamestart Paterson\surnameend},
  editor: {\sl \bibinfo{booktitle}{Automata, Languages and Programming}}, {\sl
  \bibinfo{series}{Lecture Notes in Computer Science}} \bibinfo{volume}{443},
  \bibinfo{publisher}{Springer Berlin Heidelberg}, pp.
  \bibinfo{pages}{167--180}, \doi{10.1007/BFb0032030}.

\bibitemdeclare{article}{Milner1992}
\bibitem{Milner1992}
\bibinfo{author}{Robin \surnamestart Milner\surnameend},
  \bibinfo{author}{Joachim \surnamestart Parrow\surnameend} \&
  \bibinfo{author}{David \surnamestart Walker\surnameend}
  (\bibinfo{year}{1992}): \emph{\bibinfo{title}{{A calculus of mobile
  processes, I \& II}}}.
\newblock {\sl \bibinfo{journal}{Information and computation}}
  \bibinfo{volume}{100}(\bibinfo{number}{1}), pp. \bibinfo{pages}{1--77},
  \doi{10.1016/0890-5401(92)90008-4}.

\bibitemdeclare{phdthesis}{Pet62}
\bibitem{Pet62}
\bibinfo{author}{C.~A. \surnamestart Petri\surnameend} (\bibinfo{year}{1962}):
  \emph{\bibinfo{title}{{Kommunikation mit Automaten.}}}
\newblock Ph.D. thesis, \bibinfo{school}{Bonn: Institut f{\"u}r Instrumentelle
  Mathematik, Schriften des IIM Nr. 2}.

\bibitemdeclare{inproceedings}{Phi93}
\bibitem{Phi93}
\bibinfo{author}{I.~C.~C. \surnamestart Phillips\surnameend}
  (\bibinfo{year}{1993}): \emph{\bibinfo{title}{A Note on Expressiveness of
  Process Algebra}}.
\newblock In \bibinfo{editor}{G.~L. \surnamestart Burn\surnameend},
  \bibinfo{editor}{S.~\surnamestart Gay\surnameend} \& \bibinfo{editor}{M.~D.
  \surnamestart Ryan\surnameend}, editors: {\sl \bibinfo{booktitle}{Proceedings
  of the First Imperial College Department of Computing Workshop on Theory and
  Formal Methods}}, \bibinfo{series}{Workshops in Computing},
  \bibinfo{publisher}{Springer-Verlag}, pp. \bibinfo{pages}{260--264},
  \doi{10.1007/978-1-4471-3503-6\_20}.

\bibitemdeclare{book}{SW01}
\bibitem{SW01}
\bibinfo{author}{Davide \surnamestart Sangiorgi\surnameend} \&
  \bibinfo{author}{David \surnamestart Walker\surnameend}
  (\bibinfo{year}{2001}): \emph{\bibinfo{title}{The Pi-Calculus - a theory of
  mobile processes}}.
\newblock \bibinfo{publisher}{Cambridge University Press}.

\bibitemdeclare{article}{Turing1936}
\bibitem{Turing1936}
\bibinfo{author}{Alan~Mathison \surnamestart Turing\surnameend}
  (\bibinfo{year}{1936}): \emph{\bibinfo{title}{{On computable numbers, with an
  application to the Entscheidungsproblem}}}.
\newblock {\sl \bibinfo{journal}{J. of Math}} \bibinfo{volume}{58}, pp.
  \bibinfo{pages}{345--363}.

\end{thebibliography}

\iceonly{\input{Appendix}}
\end{document}